\documentclass{aastex631}
\shorttitle{The Sun-as-a-star H$\alpha$ spectrum of solar flare}
\shortauthors{Namekata et al.}

\begin{document}

\title{
Sun-as-a-star Analysis of H$\alpha$ Spectra of a Solar Flare Observed by SMART/SDDI: Time Evolution of Red Asymmetry and Line Broadening
}

\author[0000-0002-1297-9485]{Kosuke Namekata}
\affiliation{ALMA Project, NAOJ, NINS, Osawa, Mitaka, Tokyo, 181-8588, Japan}

\author{Kiyoshi Ichimoto}
\affiliation{Astronomical Observatory, Kyoto University, Sakyo, Kyoto 606-8502, Japan}

\author{Takako T. Ishii}
\affiliation{Astronomical Observatory, Kyoto University, Sakyo, Kyoto 606-8502, Japan}


\author{Kazunari Shibata}
\affiliation{Kwasan Observatory, Kyoto University, Yamashina, Kyoto 607-8471, Japan}
\affiliation{School of Science and Engineering, Doshisha University, Kyotanabe, Kyoto 610-0321, Japan.}





\begin{abstract}

Stellar flares sometimes show red/blue asymmetries of H$\alpha$ line, which can indicate chromospheric dynamics and prominence activations. 
However, the origin of asymmetries is not completely understood.  
For a deeper understanding of stellar data, we performed a Sun-as-a-star analysis of H$\alpha$ line profiles of an M4.2-class solar flare showing dominant emissions from flare ribbons by using the data of the Solar Dynamics Doppler Imager onboard the Solar Magnetic Activity Research Telescope at Hida Observatory. 
The Sun-as-a-star H$\alpha$ spectra of the flare show red asymmetry of up to $\sim$95 km s$^{-1}$ and line broadening of up to $\sim$7.5 {\AA}.
The Sun-as-a-star H$\alpha$ profiles are consistent with spectra from flare regions with weak intensity, but they take smaller redshift velocities and line widths by a factor of $\sim$2 than those with strong intensity. 
The redshift velocities, as well as line widths, peak out and decay more rapidly than the H$\alpha$ equivalent widths, which is consistent with chromospheric condensation model and spatially-resolved flare spectra. 
This suggests that as a result of superposition, the nature of chromospheric condensation is observable even from stellar flare spectra.
The time evolution of redshift velocities is found to be similar to that of luminosities of near-ultraviolet rays (1600 {\AA}), while the time evolution of line broadening is similar to that of optical white lights.
These H$\alpha$ spectral behaviors in Sun-as-a-star view could be helpful to distinguish whether the origin of H$\alpha$ red asymmetry of stellar flares is a flare ribbon or other phenomena.

\end{abstract}

\keywords{}

\section{Introduction}

Solar and stellar flares are sudden brightenings on the surface. It is believed that flares occur through the magnetic reconnection in the coronae where the magnetic energy stored around spots is converted to kinetic, thermal, and non-thermal particle energies \citep[see,][and reference therein]{1981sfmh.book.....P,2011LRSP....8....6S}.
The released energy is transported to the chromosphere/photosphere via non-thermal particles, thermal conduction, \textcolor{black}{or possibly Alfv\'{e}n waves \citep{2011LRSP....8....6S,2008ApJ...675.1645F}}. 
The energy injection to lower atmosphere causes upward chromospheric evaporations into coronal magnetic loops (referred to as ``flare loops") and downward chromospheric condensations in flare foot points (referred to as ``flare ribbons"). 
Also, solar flares \textcolor{black}{may be associated with} upward filament/prominence eruptions and coronal mass ejections (CMEs). 

In the case of the Sun, the nature of major flare-related phenomena, such as (i) flare ribbons, (ii) flare loops, and (iii) filament/prominence eruptions, have been investigated in very large amount of studies.
H$\alpha$ line has been historically used as a strong tool to diagnose these phenomena. 
Flare ribbons mostly show redshifted, broadened H$\alpha$ emission profiles, which are thought to represent chromospheric condensation moving downward \citep[e.g.,][]{1984SoPh...93..105I,1985ApJ...289..414F,2012PASJ...64...20A}.
Only several percent of flares show a blueshifted profile in chromospheric lines \citep{1962BAICz..13...37S,1990ApJ...363..318C}, which may relate with a ``cool" evaporation flow \citep{2018PASJ...70..100T}.
Flare loops are often seen as a combination of weak emissions in H$\alpha$ line center and redshifted absorption when the solar flares occur on disk, but can be seen as significant emission when it occurs off the limb \citep[e.g.,][]{1996SoPh..166...89W,2016NatSR...624319J}. 
Eruptive phenomena can be mostly seen blue/redshifted absorption on the solar disk (filament eruption), but they appears in emission outside the disk (prominence eruption) \citep{2014LRSP...11....1P}.

In recent years, there has been a growing interest in stellar flare observations, especially in terms of the exoplanet habitability \citep[e.g.,][]{2010AsBio..10..751S,2016NatGe...9..452A,2020IJAsB..19..136A,2019ApJ...881..114Y} and as a proxy of possible extreme flares (``superflares") on our Sun \citep[e.g.,][]{2012Natur.485..478M,2012JGRA..117.8103S,2013ApJS..209....5S,2013PASJ...65...49S,2012JGRA..117.8103S,2019ApJ...876...58N,2020arXiv201102117O}. 
The intense radiations and CMEs associated with stellar flares can have a severe impact on the planetary environment \citep[e.g.,][]{2021LRSP...18....4T}, and there is an urgent need to clarify their characteristics.
In the case of stellar flares, as in the case of solar flares, spectroscopic observations of H$\alpha$ line have long been used for characterizing the cool plasma conditions, and many studies have reported the line broadening \citep[e.g.,][]{2017ApJ...837..125K,2020PASJ...72...68N} and the blue/redshifted H$\alpha$ profiles during stellar flares \citep[e.g.,][]{1993A&A...274..245H,2016A&A...590A..11V,2018PASJ...70...62H,2019A&A...623A..49V,2020arXiv201200786K,2020PASJ..tmp..253M}.
The major difficulty in stellar flare observations is to distinguish between the major phenomena (i)$\sim$(iii) from the observed H$\alpha$ spectra since they are not spatially resolved, unlike solar flares.
For example, on M, K-dwarf flares, red asymmetry\footnote{Note that the red/blue ``asymmetry" means the red/blue wind enhancements, not absorptions.} is interpreted not only as chromospheric condensation \citep{1993A&A...274..245H}, but also as downward motion in flare loops or backward prominence eruptions \citep[e.g.,][]{2020arXiv201200786K}.
Blue asymmetry is interpreted as cool upward flow associated with evaporation or prominence eruptions \citep[e.g.,][]{1990A&A...238..249H,1994A&A...285..489G,2004A&A...420.1079F,2011A&A...534A.133F,2016A&A...590A..11V,2018PASJ...70...62H,2019A&A...623A..49V,2020MNRAS.499.5047M,2020PASJ..tmp..253M}.

Solar flares can provide basic knowledge to distinguish those phenomena in stellar observations, not only because the Sun can be observed in a spatially-resolved manner, but also because it can be regarded as the nearest star.
Observations and analyses of solar flares in ``Sun-as-a-star" view (i.e., spatial integration) has been thought to be of great use for a deeper understanding of the observed data of stellar flares.
In fact, Sun-as-a-star observations of flares in X-rays \citep{1997A&A...325..782R,2002ApJ...577..422S,2017PASJ...69....7N,2021NatAs...5..697V}, extreme-ultraviolet rays \citep{2016SoPh..291.1761H,2021NatAs...5..697V}, and white lights \citep{2010NatPh...6..690K,2011A&A...530A..84K,2017ApJ...851...91N} have made a significant contribution to the interpretation of stellar flare observations.
However, Sun-as-a-star analyses of H$\alpha$ spectrum of solar flares have been seldom reported before.
Although we know what H$\alpha$ spectrum of each phenomenon looks like in the spatially-resolved view, we do not know which phenomenon among is contributing to the change in the total Sun-as-a-star spectral variation, to what extent, and how.

The Solar Dynamics Doppler Imager (SDDI) is an instrument installed on the Solar Magnetic Activity Research Telescope (SMART) at Hida Observatory \citep{2017SoPh..292...63I}.
It has a unique ability to take full disk solar images at every \textcolor{black}{0.25 {\AA}} from the H$\alpha$ line center $-$9.0 {\AA} to the H$\alpha$ line center +9.0 {\AA} with 12-20 seconds time cadence, enabling us to make the Sun-as-a-star H$\alpha$ spectrum.
By using the data of SDDI, our previous study performed the Sun-as-a-star analysis of two C-class solar flares associated with a filament eruption and a surge \citep{Namekata2020Sci}.
The Sun-as-a-star spectra shows a blueshifted absorption due to the eruptive filaments with gradually decaying velocity.
We compared them with H$\alpha$ spectrum of a superflare on a solar-type star, and surprisingly found that the stellar data is very similar to the Sun-as-a-star filament eruption events.
This comparison has become a clear evidence that on the solar-type star, a filament eruption has happened, which is similar to solar phenomena.

Since a wide variety of H$\alpha$ spectral changes of stellar flares have been reported in recent years, it is necessary to perform Sun-as-a-star analysis of H$\alpha$ line on a wider variety of solar phenomena in order to connect the solar observations to stellar observations. 
Also, H$\alpha$ stellar flares are often observed simultaneously with other wavelengths, such as white-light emission with \textit{TESS} observations \citep[e.g.,][]{2020PASJ..tmp..253M,Namekata2020Sci,2022arXiv220109416N}, so simultaneous Sun-as-a-star observations with other wavelength (e.g., white-light, ultraviolet, and X-ray) can provide a key information for further solar-stellar connection.
In this paper, we report a Sun-as-a-star analysis of H$\alpha$ spectra of an M4.2-class solar flare occurred at 01:05 UT on 2017 September 5.
In the event, the H$\alpha$ radiation from flare ribbons occurred near the disk center and was dominant compared to those of the other component of flares, such as post-flare loop and filament eruption.
Moreover, it showed strong redshift spectra and white-light emission; therefore it is a very good target to investigate the behavior of flare ribbons in relation with other wavelengths in the Sun-as-a-star view. 
Our aims of this study are the following three points: 
(1) to reformulate the Sun-as-a-star analysis method of solar flares using SMART/SDDI data in more detail than \cite{Namekata2020Sci}, 
(2) to investigate the nature of red asymmetry and line broadenings of the Sun-as-a-star H$\alpha$ spectrum in comparison with other wavelengths,
and (3) to investigate the relation of the H$\alpha$ spectra of a solar flare in the Sun-as-a-star view and those in the spatially-resolved view.
We introduce the dataset and Sun-as-a-star analysis method in Section \ref{sec:2}, show the results in Section \ref{sec:3}, discuss the results in Section \ref{sec:4}, and describe summary and future work in Section \ref{sec:5}.

\section{Dataset and Analysis Method}\label{sec:2}

\subsection{M4.2-class solar flare on 2017 September 5}\label{sec:2-1}

\begin{figure}
\epsscale{0.5}
\plotone{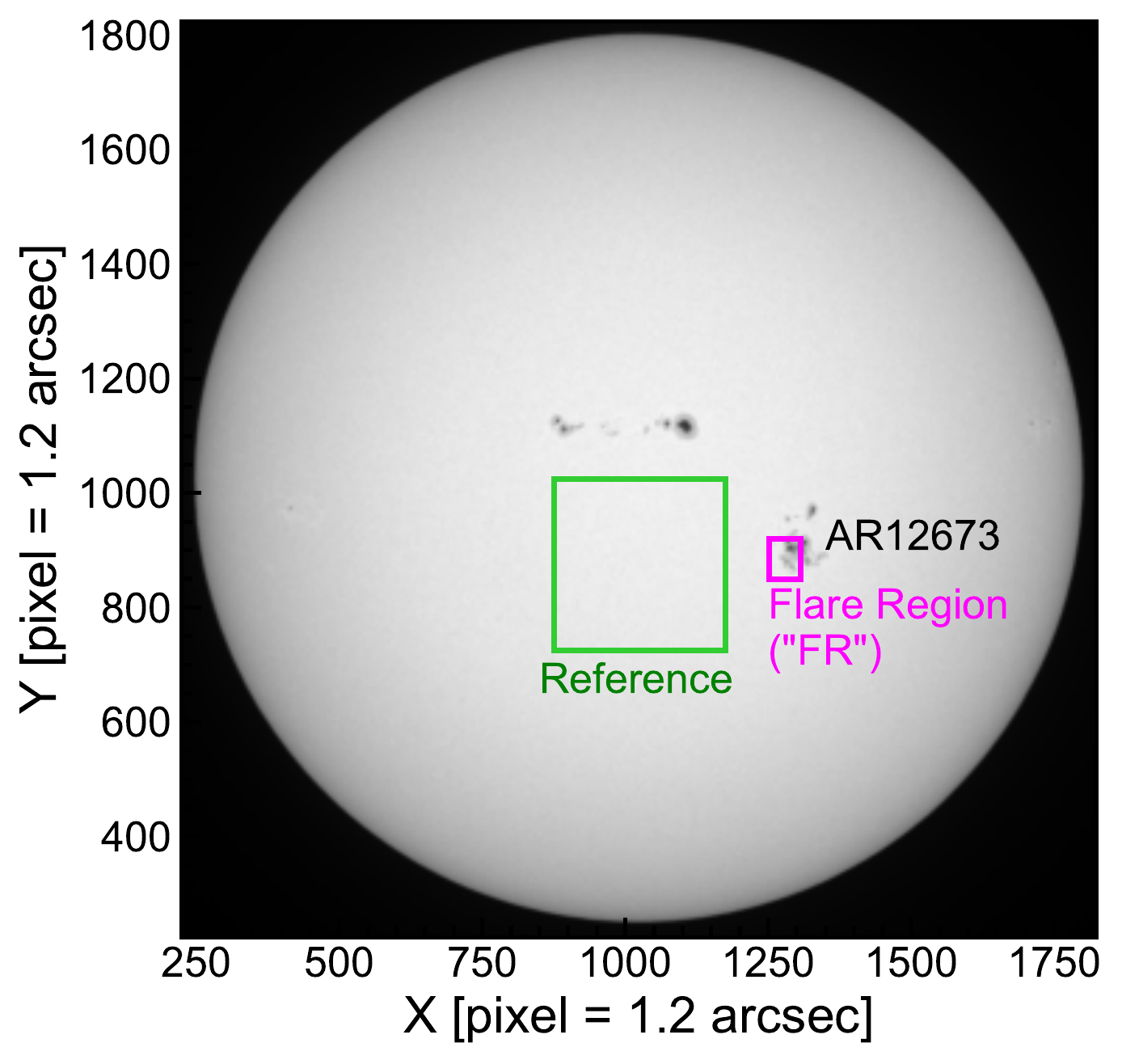}
\caption{A \textcolor{black}{full-disk} solar image at H$\alpha$ far wing (6553.8 {\AA}) observed by SMART/SDDI on 2017 September 5. \textcolor{black}{The magenta square indicates the} field of view of Figure \ref{fig:2} (referred to as Flare Region or ``FR" in the text), located around the flare-productive active region 12673 (AR12673). \textcolor{black}{The green square indicates} a reference region (``ref." in the text).
Note that this image is rotated 22.5 degrees counterclockwise to the North pole of the Sun and upside is the celestial North.}
\label{fig:1}
\end{figure}

We analyzed an M4.2-class solar flare which occurred in active region 12673 (AR 12673) at 01:05 UT on 2017 September 5.
AR 12673 is known as the most flare-productive AR in solar cycle 24, producing many M- and X-class flares \citep{2018ApJ...856...79Y,2018ApJ...861...28S,2018SpWea..16.1549C,2020RAA....20...23S,2021ApJ...908..132Y}.
Our target solar flare occurred when the AR 12673 was not far from the disk center of the Sun as in Figure \ref{fig:1}.
The detailed H$\alpha$ map is shown in Figure \ref{fig:2} (see, Section \ref{sec:2-2}) and its magnetic field map is shown in Figure \ref{fig:3} (see, Section \ref{sec:2-3}).
We selected this event for the following reasons: (i) It was observed with the SMART telescope at Hida Observatory under good weather condition. (ii) The event occurred near the disk center. (iii) The flare emission is dominant, and the associated eruption phenomena and coronal rain are weak.
Because the event satisfied the above criteria, we consider the event as a favorable solar flare to understand the nature of the flare ribbon in the Sun-as-a-star view.

\subsection{SMART/SDDI data analysis}\label{sec:2-2}

\begin{figure}
\epsscale{1}
\plotone{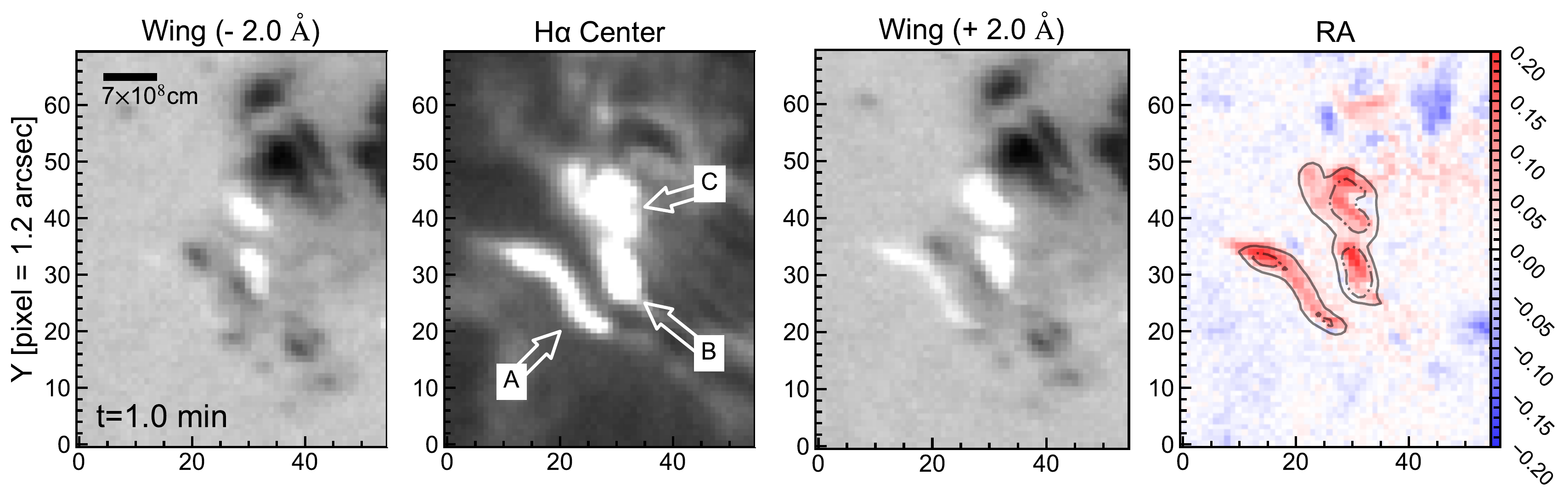}
\plotone{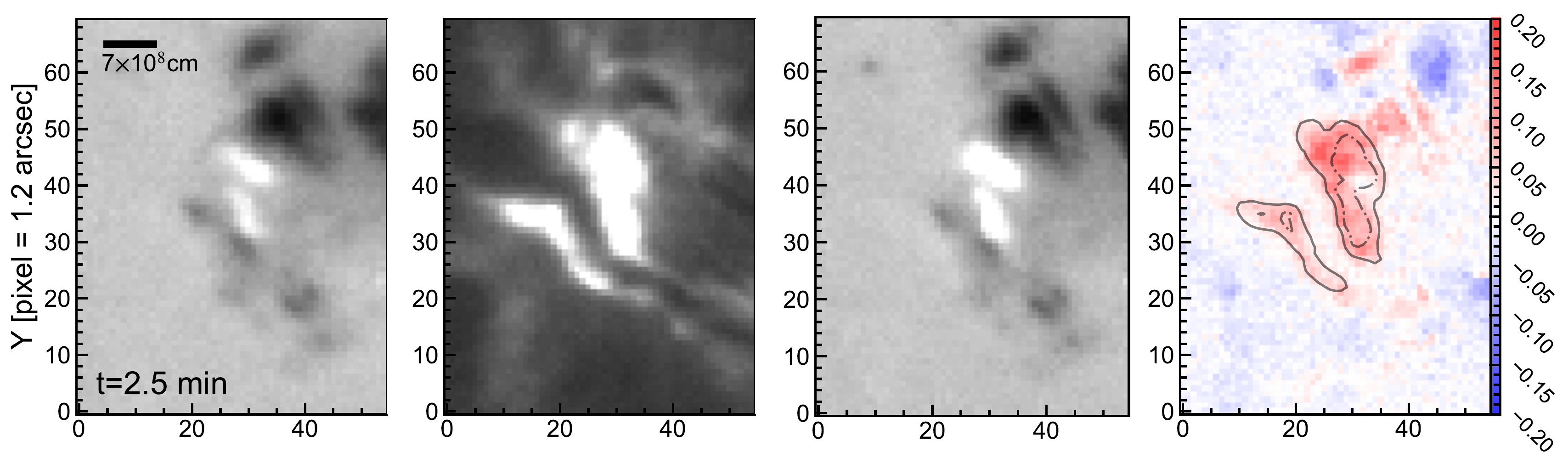}
\plotone{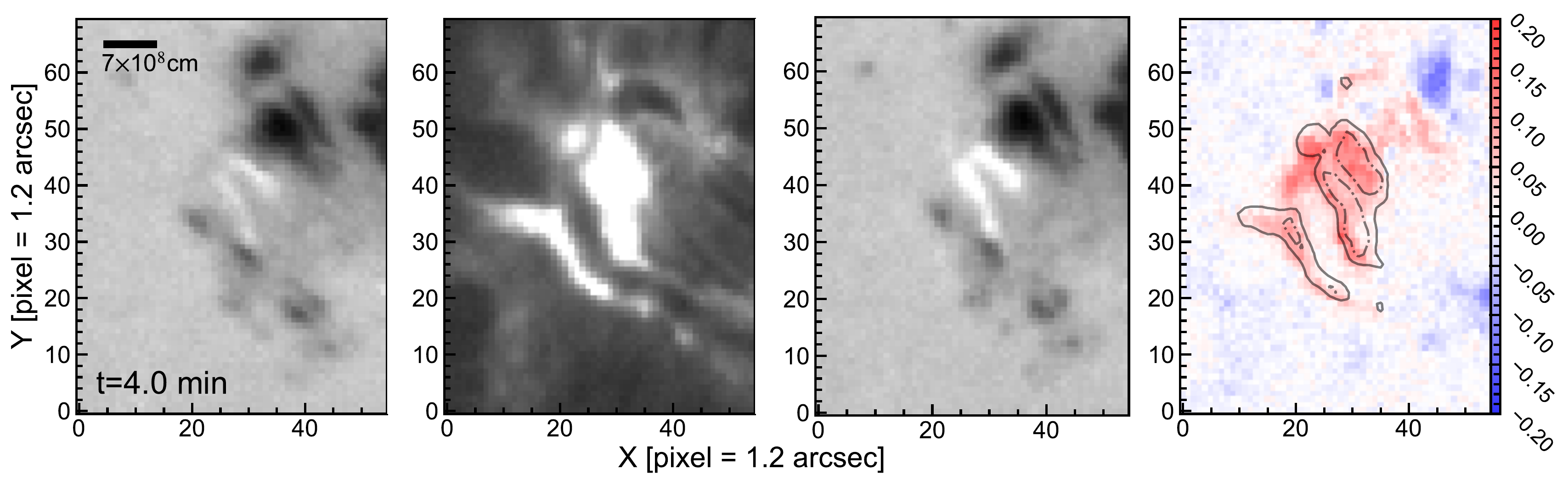}
\caption{SMART/SDDI images of the M4.2-class solar flare on 2017 September 5. 
The field of view of each panel \textcolor{black}{corresponds} to FR in Figure \ref{fig:1}.
From top to bottom, images at the time of 1 minute, 2 minutes, and 4  minutes after the start of the flare (01:05:35 UT) are shown.
From left to right, images of $\lambda_{\rm cen}$$-$2{\AA}, $\lambda_{\rm cen}$, $\lambda_{\rm cen}$+2{\AA}, and red asymmetry RA (Equation \ref{eq:ra}) are shown, where $\lambda_{\rm cen}$ is the H$\alpha$ center wavelength.
\textcolor{black}{In the top of H$\alpha$ center image, the three flare ribbons A, B, and C are labeled with the arrows (see Figure \ref{fig:3} for the counterparts of UV flare ribbons)}.
}
\label{fig:2}
\end{figure}

Here we introduce the Sun-as-a-star analysis method of H$\alpha$ flare spectrum taken by SMART/SDDI in more detail than \cite{Namekata2020Sci}.
SDDI has been conducting a monitoring observation of the Sun since 2016. 
It takes the full-disk Sun images at 73 wavelength points at every \textcolor{black}{0.25 {\AA}} from $\lambda_{\rm cen}$$-$9.0 {\AA} to $\lambda_{\rm cen}$+9.0 {\AA}, where $\lambda_{\rm cen}$ is the H$\alpha$ line center wavelength.
For this events, each set of images is observed with a time cadence of 15 seconds and a pixel size of 1.2 arcsec.
Figure \ref{fig:1} is the full disk image of SDDI at 6553.8 {\AA}.
Figure \ref{fig:2} shows the partial images of the M4.2-class solar flare on 2017 September 5 for $\lambda_{\rm cen}$-- 2 {\AA},  $\lambda_{\rm cen}$, and $\lambda_{\rm cen}$ + 2 {\AA}. 
The rightmost panels show the red asymmetry ``RA", which is defined as  
\begin{eqnarray}\label{eq:ra}
RA = \frac{I({\lambda_{\rm cen}+1\,{\rm \AA}}, t, x, y) - I({\lambda_{\rm cen}-1\,{\rm \AA}}, t, x, y)}{I({\lambda_{\rm cen}+1\,{\rm \AA}}, t, x, y) + I({\lambda_{\rm cen}-1\,{\rm \AA}}, t, x, y)},
\end{eqnarray}
where $I(\lambda, t, x, y)$ is the intensity at the position ($x$,$y$), a wavelength $\lambda$, and time $t$ \citep[cf.,][]{2012PASJ...64...20A}.

In principle, the ``real" Sun-as-a-star spectrum will be obtained by simply performing a full-disk integration at each wavelength. 
However, in the case of the full-disk integration, we could not obtain a clear spectral variation of the M4.2-class solar flare. 
This could be not only due to intrinsic brightness variation of the solar surface, but also  due to noise in spectral profiles induced by atmospheric fluctuations and weather variations since each wavelength are not exactly observed at the same time (15 sec lag between $-$9 {\AA} and $+$9 {\AA}).
Therefore, in our study, we spatially integrate the intensity inside a partial region of the SDDI images that are large enough to cover the visible phenomena (the magenta region is Figure \ref{fig:1}, hereafter referred to as Flare Region ``FR").
This is not exactly a Sun-as-a-star analysis, but it can be virtually treated as a Sun-as-a-star by finally normalizing it with the solar luminosity \citep[also referred to as ``virtual" Sun-as-a-star in][]{Namekata2020Sci}.

In the following, we will introduce the method of this ``virtual" Sun-as-a-star analysis: 
First, we performed a pixel-level position correction for every image using non-flaring spots in AR 12673 as a marker in order to correct position changes due to solar rotation and seeing fluctuation.
Before going to the spatial integration, each image was normalized by the value of median of the reference region (the green region in Figure \ref{fig:1}) for each wavelength in order to cancel the variation due to the weather condition and solar altitude,
\begin{eqnarray}
I'(\lambda, t, x, y) = \frac{I(\lambda, t, x, y)}{median_{x,y \in \rm ref.}({I}(\lambda, t, x, y))},
\label{eq:0p}
\end{eqnarray}
where ``ref." means the reference region.
Here, $I'(\lambda, t, x, y)$ is 2D-image normalized by the median value of ``each wavelength" of the reference region. 
Then, the 2D-image spectrum normalized by the median value of ``continuum level" of the reference region $I_{\rm n}$($\lambda$, $t$, $x$, $y$) can be obtained as 
\begin{eqnarray}
I_{\rm n}(\lambda, t, x, y) &=& I'(\lambda, t, x, y){\rm F_{\rm n}}(\lambda)\\
 &=& \frac{I(\lambda, t, x, y)}{median_{x,y \in \rm ref.}({I}(\lambda, t, x, y))}{\rm F_{\rm n}}(\lambda),
\label{eq:0}
\end{eqnarray}
where F$_{\rm n}$($\lambda$) is a template solar quiescent H$\alpha$ spectrum which is convolved with SDDI instrumental profile and normalized by the continuum level \citep{2017ApJ...843L..24S}.  
Here, we define a spectrum normalized by the median value of continuum level of the reference region as  $L(\lambda, t, {\rm region})$.
We integrated the intensity inside the Flare Region ``FR"  (the magenta region in Figure \ref{fig:1}) to obtain the spatially-averaged H$\alpha$ spectrum $L(\lambda, t, {\rm FR})$ as
\begin{eqnarray}
L(\lambda, t, {\rm FR}) = \sum_{x,y \in \rm FR} {I_{\rm n}}(\lambda, t, x, y).
\label{eq:1}
\end{eqnarray}
In the same manner, the ``real" Sun-as-a-star spectra (normalized by the median of the reference region continuum) can be obtained as
\begin{eqnarray}
L(\lambda, t, {\rm full \, disk}) = \sum_{x,y \in {\rm full \, disk}} {I_{\rm n}}(\lambda, t, x, y),
\label{eq:2}
\end{eqnarray}
Again, it is difficult to measure a solar flare component with Equation \ref{eq:2} as mentioned above, so this full disk spectrum is used just to scale the Equation \ref{eq:1} to Sun-as-a-star.
The correction factor to convert Equation \ref{eq:1} to the full disk continuum unit can be obtained as $L(\lambda_{\rm cont}, t_0, {\rm full \, disk})$, where $\lambda_{\rm cont}$ = $\lambda_{\rm cen}$ $-$ 9 {\AA} is a wavelength of the continuum level, $\lambda_{\rm cen}$ is the H$\alpha$ center wavelength, and $t_0$ is a time in pre-flare phase.
Here, a pre-flare-subtracted flaring spectrum at the time $t$ (normalized by the median of the reference region) can be expressed as $L(\lambda, t, {\rm FR}) - L(\lambda, t_0, {\rm FR})$ ($\approx$ $L(\lambda, t, {\rm full \, disk}) - L(\lambda, t_0, {\rm full \, disk})$).
Therefore, a pre-flare-subtracted flaring spectrum normalized by the full disk continuum level, $\Delta S_(\lambda, t)$, can be expressed as
\begin{eqnarray}\label{eq:3}
\Delta S(\lambda, t) &=& \frac{L(\lambda, t, {\rm FR}) - L(\lambda, t_0, {\rm FR})}{L(\lambda_{\rm cont}, t_0, {\rm full \, disk})}\\
\Biggl( &\approx& \frac{L(\lambda, t, {\rm full \, disk}) - L(\lambda, t_0, {\rm full \, disk})}{L(\lambda_{\rm cont}, t_0, {\rm full \, disk})} \Biggr) .
\end{eqnarray}
If we add a full disk spectrum (normalized by full disk continuum) to Equation \ref{eq:3}, we can obtain the Sun-as-a-star spectrum of the H$\alpha$. 
However, the variations of flare is too small compared to the original solar full disk spectrum (an order of $\sim$10$^{-4}$).
Therefore, we show Equation \ref{eq:3} as a flaring spectrum in this study, and hereafter, we sometimes call $\Delta S(\lambda, t)$ a Sun-as-a-star (flare) spectrum by omitting ``pre-flare-subtracted" for simplicity.
Note that the analysis method described above is the same as those in \cite{Namekata2020Sci}\footnote{Note that Equation 1 in \cite{Namekata2020Sci} has a typo, and $t_0$ is a correct expression in its denominator. We correct it here.}, but the expression of equations has been a bit improved in this paper.
We calculated the equivalent width EW = $\int \Delta S(\lambda, t) d\lambda$ in the unit of {\AA}.
We also calculated the H$\alpha$ flare radiated energy by simply multiplying the EW by solar contiuum luminosity around H$\alpha$ \citep[cf.,][]{Namekata2020Sci}.

\subsection{SDO/HMI and AIA data analysis}\label{sec:2-3}

\begin{figure}
\epsscale{1}
\plotone{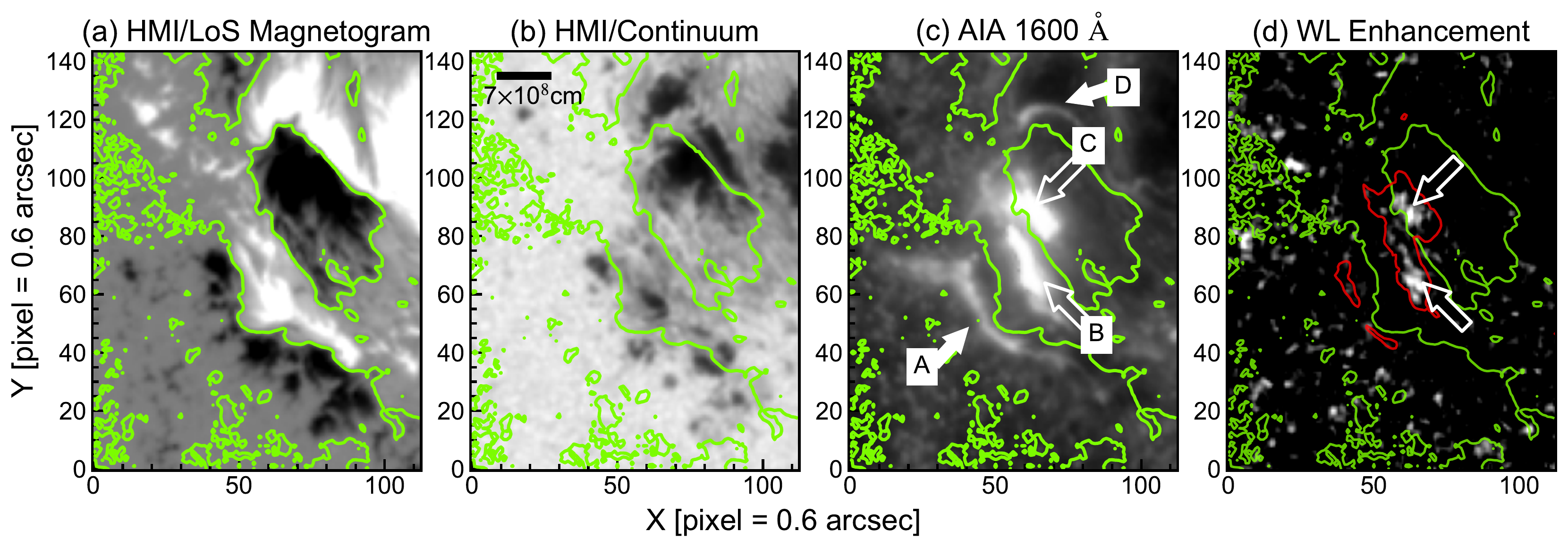}
\caption{(a) SDO/HMI light-of-sight magnetogram of the Flare Region ``FR" around AR12673 at 01:00:34 UT ($t$ = 5.0 minutes) on 2017 September 5. The green contour represents the polarity inversion line. (b) SDO/HMI continuum at 01:08:05 UT ($t$ = 2.5 minutes). (c) SDO/AIA 1600 {\AA} intensity at 01:08:14 UT ($t$ = 2.6  minutes). The four flare ribbons A, B, C, and D are labeled with the arrows. (d) Pre-flare-subtracted SDO/HMI continuum image at 01:08:05 UT ($t$ = 2.5 minutes). The white color is the white-light emission. The red contour represents the SDO/AIA 1600 {\AA} intensity of the third panel. The white-light emissions in the main flare ribbons B and C are indicated with arrows (see, Figure \ref{fig:2}).
}
\label{fig:3}
\end{figure}

In this section, we briefly explain the analysis of the data obtained by the Helioseismic and Magnetic Imager (HMI) and the Atmospheric Imaging Assembly (AIA) onboard the \textit{Solar Dynamics Observatory} ($SDO$).
The purpose of analyzing these data is to overview the magnetic field configuration in which this event occurred and to investigate how this solar flare behaved in white light and UV in the Sun-as-a-star view.

Figure \ref{fig:3} shows the magnetic field configuration \textcolor{black}{around the time of the flare (5 min after the onset)} and the intensity of AIA 1600 {\AA}. 
Four flare ribbons are observed in AIA 1600 {\AA} (A, B, C, and D in the figure). 
According to the NLFF calculation of \cite{2021ApJ...908..132Y}, the flare ribbons of A and B, B and C, and C and D are connected by magnetic field lines, and most of the flare radiation is emitted from flare ribbons B and C in  AIA 1600 {\AA}. 
Figure \ref{fig:2} shows that also in H$\alpha$, flare ribbons B and C \textcolor{black}{are} dominant, while flare ribbon A is visible.

Next, we show the Sun-as-a-star analysis of the associated white-light flare with the use of the SDO/HMI continuum \citep{2017ApJ...851...91N}.
The colormap in Figure \ref{fig:3}(d) shows the enhanced white-light flare emission (white region) with the AIA 1600 {\AA} contour (red).
We first made a time-integrated AIA 1600 {\AA} image and made a mask image where the intensity is over 10 \% of the peak intensity.
By integrating the white-light enhancement inside the mask, we have obtained the total counts of the white-light flare $\rm DN_{\rm flare}$.
After that, the white light enhancement as Sun-as-a-star ($\rm DN_{\rm flare}$/$\rm DN_{\rm \odot}$) was obtained by dividing it by the full disk count of the SDO/HMI continuum $\rm DN_{\rm \odot}$ (see Figure \ref{fig:6}).
The flare flux and energy \textcolor{black}{were} calculated in the same way as \cite{2017ApJ...851...91N} by assuming the 10,000 K blackbody radiation \citep{2010NatPh...6..690K,2011A&A...530A..84K}.

We also made a Sun-as-a-star light curve of AIA 1600 {\AA}.
The radiation source of AIA 1600 {\AA} band is the continuum and emission lines such as C IV, which originates from the transition region and upper photosphere \citep{2019ApJ...870..114S}.
Although the formation heigh of AIA 1600 {\AA} is complicated with the contribution from multiple emission lines during flares \citep[mostly from transition region via C IV lines;][]{2019ApJ...870..114S}, the rising phase of AIA 1600 {\AA} and GOES X-ray light curves shows a relation of \textit{Neupert} effect \citep{2009ApJ...694L..74N,2013ApJ...774...14Q,2018ApJ...863..124D}.
Hence, the AIA 1600 {\AA} light curve is plotted as a reference for impulsive energy release in the lower atmosphere (i.e. a proxy of non-thermal heating) in this paper (see Section \ref{sec:3-1}).

\subsection{Measurements of line shift and width}\label{sec:2-4}

\begin{figure}
\epsscale{0.5}
\plotone{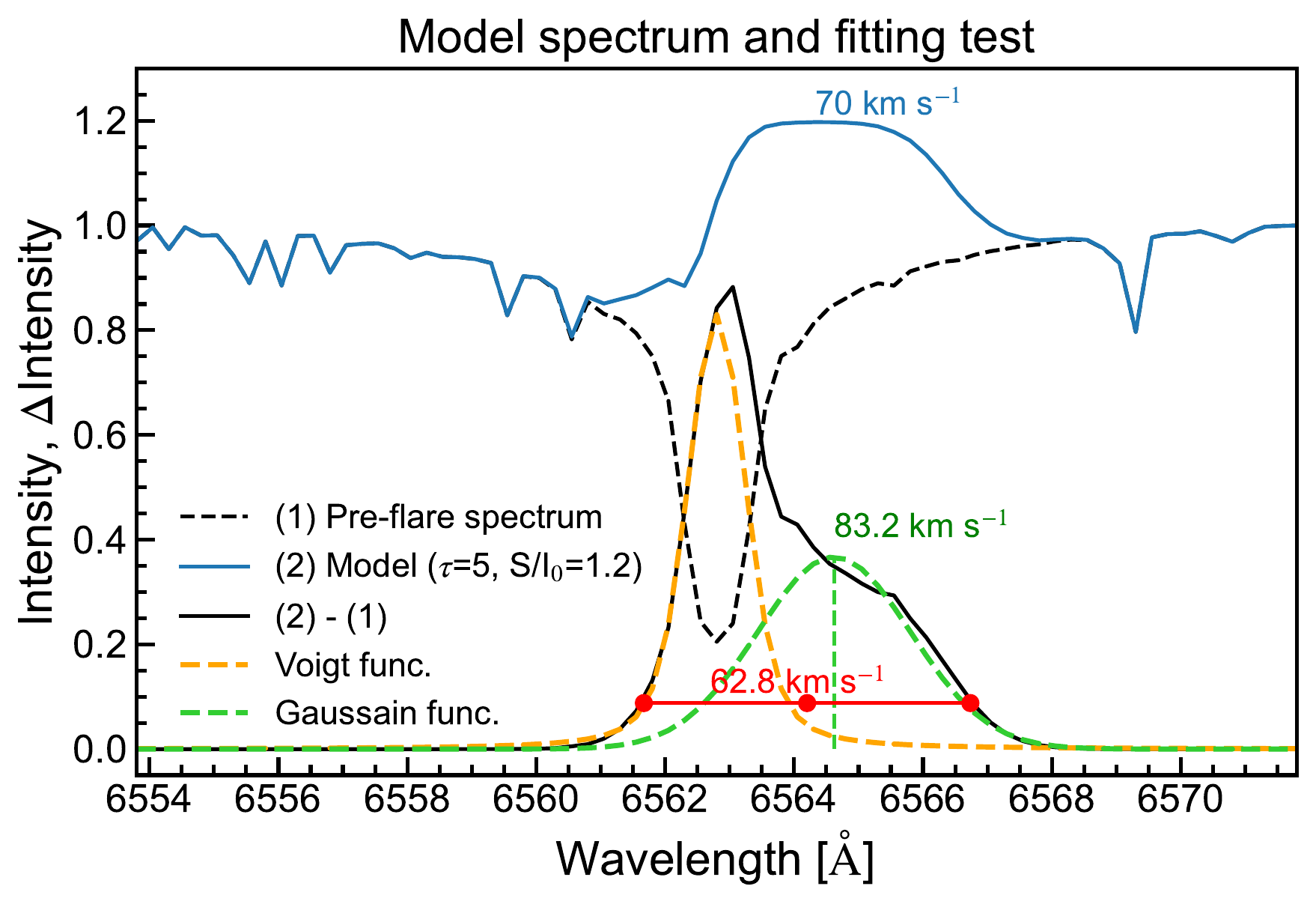}
\caption{Model H$\alpha$ spectrum for the fitting test of the two methods. A black dashed line (1)  is a background solar spectrum. A blue solid line (2) is a model flare spectrum. The model spectrum is calculated with the Becker's cloud model \citep{1964PhDT........83B} with optically thickness $\tau$ of 5, velocity of 70 km s$^{-1}$, and source function $S/I_0$ of 1.2. 
A black solid line is a pre-flare-subtracted spectrum, i.e., (2)$-$(1). 
Orange and green lines represent the result of the 2 component fitting with central Voigt function and redshifted Gaussian function, respectively. 
Red points correspond to the result of the single wing component fitting, where each point represents the left wing wavelength, redshifted wavelength, and right wing wavelength. 
}
\label{fig:4}
\end{figure}

We have employed two fitting methods to measure the line shift and line width of H$\alpha$ flaring emission using the pre-flare-subtracted spectra, $\Delta S(\lambda,t)$, defined in Section \ref{sec:2-2}: (1) a two component fitting method (refer to as ``2 Component" in each figure) and (2) a single wing component fitting (refer to as ``Wing Component" in each figure).
The method (1) is often used in stellar flare observations \citep{2019A&A...623A..49V,2020PASJ..tmp..253M,2022ApJ...928..180W}, and the method (2) is similar to that used in solar flare observations \citep[cf.,][]{1984SoPh...93..105I}. 
The method (1) is the following procedure (see Figure \ref{fig:4} as example): First, the data on the short wavelength side of the H$\alpha$ center is fitted with the Voigt function to obtain the H$\alpha$ center component (orange line in Figure \ref{fig:4}). At this time, the line width of the Voigt function is determined. Then, the remaining redside excess component is fitted with a Gaussian function to obtain its redshift velocity and intensity (green line in Figure \ref{fig:4}).
The line width is determined as widths of one tenths of peak intensity for the central Voigt function.
The method (2) is the following procedure: Only the line wing component is fitted with a single-velocity Voigt function, and its line width (one tenths of peak intensity level) and shifted velocity are obtained. 
The wing criterion used for the fit is $<$0.2 of continuum level for the Sun-as-a-star data and $<$0.25 of continuum level for the spatially resolved data. 
The line width is simply determined as widths of one tenths of peak intensity of spectrum.

Figure \ref{fig:4} shows the test for the fitting methods for a modeled pre-flare-subtracted spectrum (a black solid line), which is calculated with the Becker's cloud model \citep{1964PhDT........83B} with optically thickness $\tau$ of 5, velocity of 70 km s$^{-1}$, and source function $S/I_0$ of 1.2 (original spectrum is indicated with blue line). 
\textcolor{black}{We applied the above fitting methods to the pre-flare-subtracted spectra with the aim of an application to stellar flares. In the case of the Sun where the background H$\alpha$ spectrum is usually absorption, the pre-flare-subtracted spectrum is usually shaped as a sum of the shape of flare spectrum (blue line) and that of the inverse of background spectrum (black dashed line). This results in an apparent enhancement of the central component in the pre-flare-subtracted spectrum (a black solid line), but would not significantly affect the estimation of shifted and/or broadened component.} 
For this modeled spectrum in Figure \ref{fig:4}, the estimated velocity was 83.2 km s$^{-1}$ and 62.8  km s$^{-1}$ with method (1) and (2), respectively.
After the tests for several cases, we noticed that both methods have their own advantages and disadvantages as follows: 
The method (1) is a good fitting method when the flaring atmosphere has both stationary and redshift components \citep[e.g., seen in numerical simulations;][]{2015SoPh..290.3487K}.
\textcolor{black}{However, it should be noted that the subtraction of the stationary component can improperly remove the blue wing of the red-shift component.
Therefore, the method can overestimate the velocity due to the overfitting of the central component when flaring atmosphere is a single-velocity atmosphere and/or when the velocity is low.}
The method (2) gives correct solution if both wings of Ha emission are produced by the redshift component, but underestimates the velocity in the case of multiple atmospheric components.
We have tested how the methods (1) and (2) can estimate the input velocity with a simple optically-thick H$\alpha$ flaring spectrum for several cases, but it is difficult to say which method is better to obtain the real velocity as in Figure \ref{fig:4}. Hence, we have applied both methods in this paper (see, Section \ref{sec:3-2}, \ref{sec:3-3}, and \ref{sec:3-4}).

\section{Result}\label{sec:3}

\subsection{Sun-as-a-star H$\alpha$ spectra and multi-wavelength light curves}\label{sec:3-1}

\begin{figure}
\epsscale{0.5}
\plotone{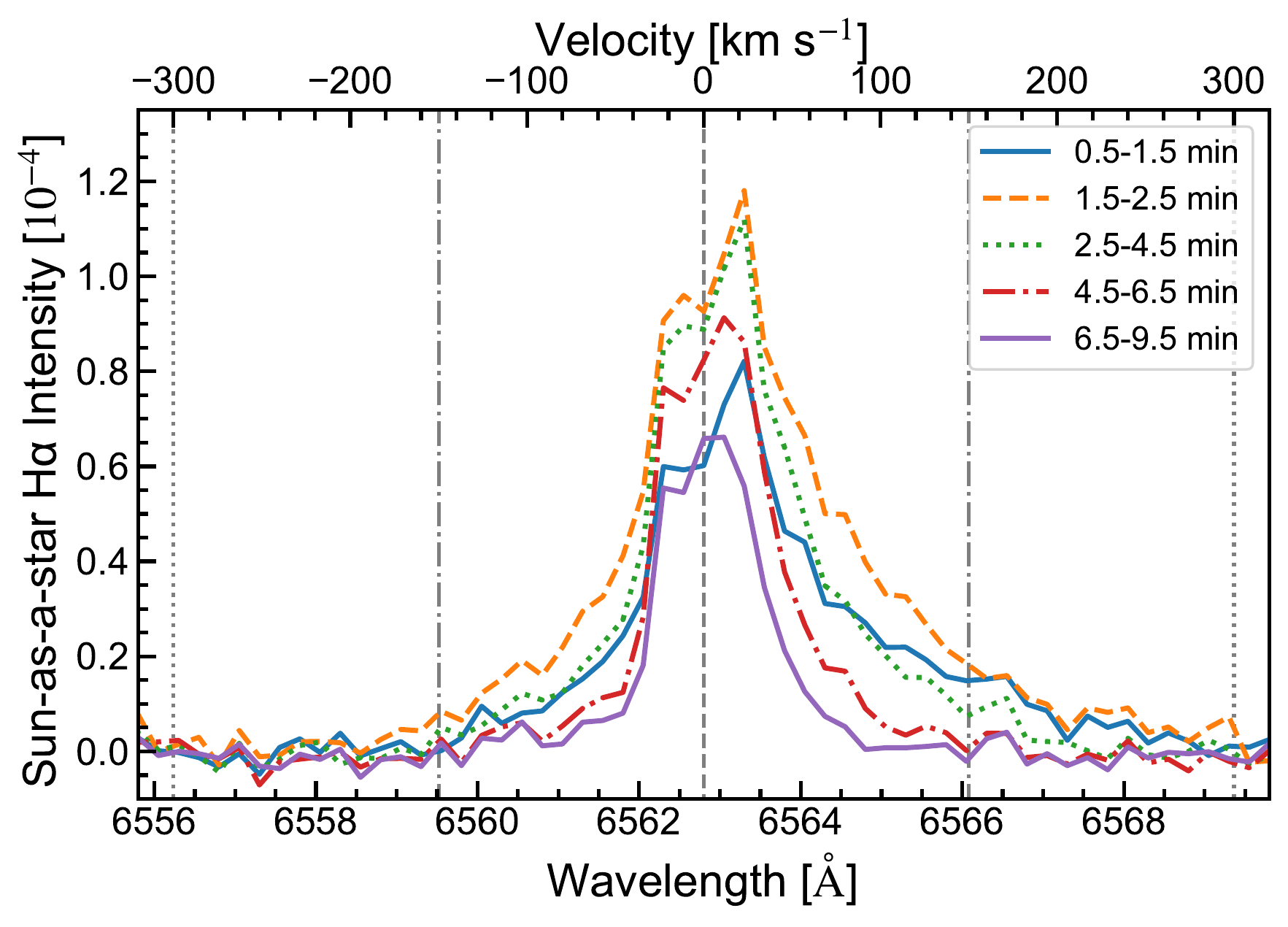}
\epsscale{0.5}
\plotone{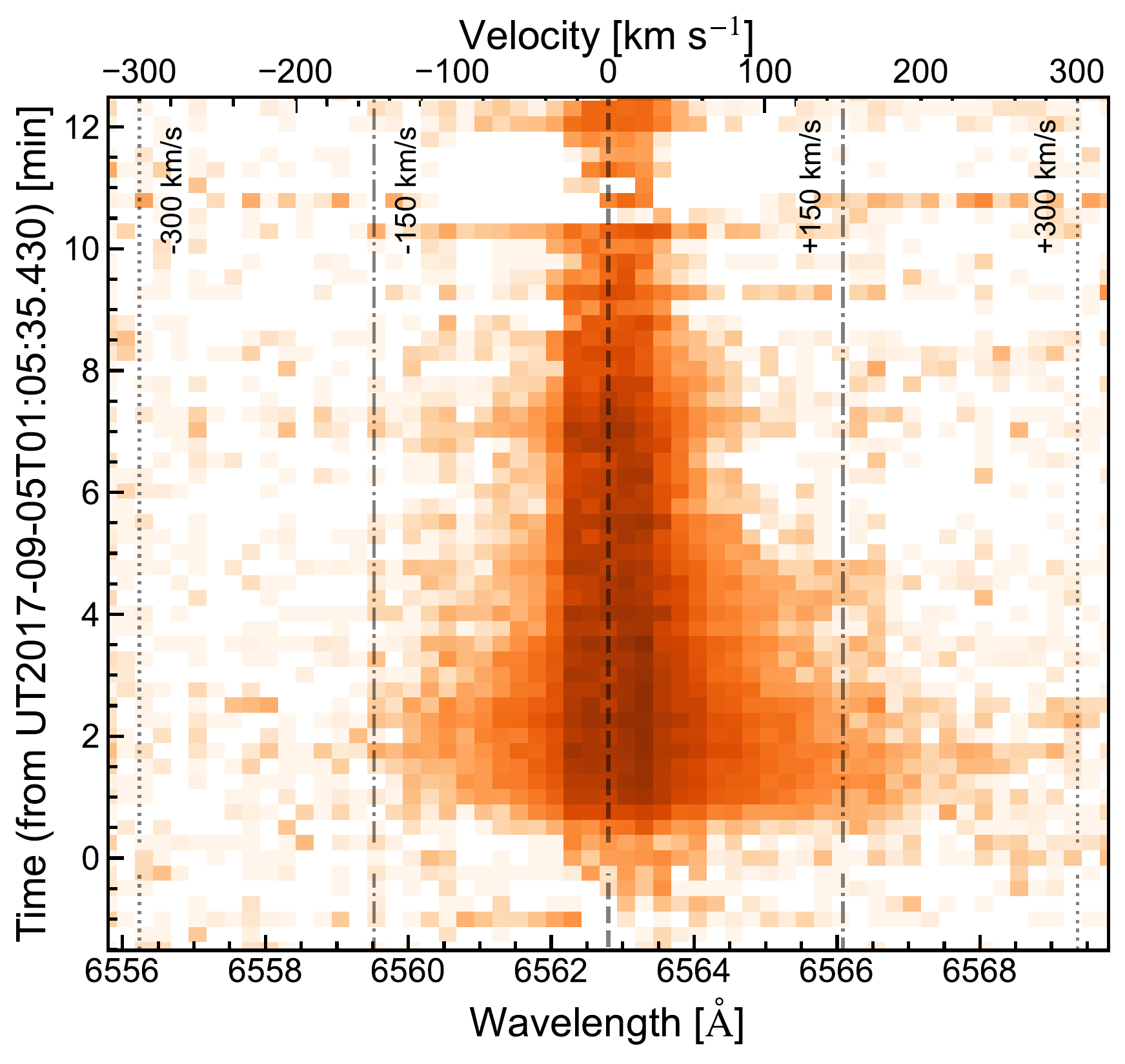}
\caption{Sun-as-a-star H$\alpha$ spectra of the solar flare. (left) The spectra in the each time bin indicated in the legend. (right) The time evolution of the Sun-as-a-star H$\alpha$ spectra in the time-wavelength plane. 
}
\label{fig:5}
\end{figure}

Figure \ref{fig:5} shows the time evolution of the Sun-as-a-star H$\alpha$ spectrum $\Delta S(\lambda, t)$ for the flare of interest obtained by the SDDI.
We found that the spectra show the enhanced red wing enhancement even in the Sun-as-a-star view, in most of the periods.
This is the same as the previous studies performing spatially-resolved analyses \citep[e.g.,][]{1984SoPh...93..105I}.
In addition, the Sun-as-a-star spectrum shows the significant line broadenings during the flare.
Qualitatively, both the redshift and broadening are prominent especially in early 0.5$\sim$4.5 minute (i.e., the impulsive phase), but as the flare decays, they become gradually weaker (see Section \ref{sec:3-2} for quantitative results).

\begin{figure}
\epsscale{0.54}
\plotone{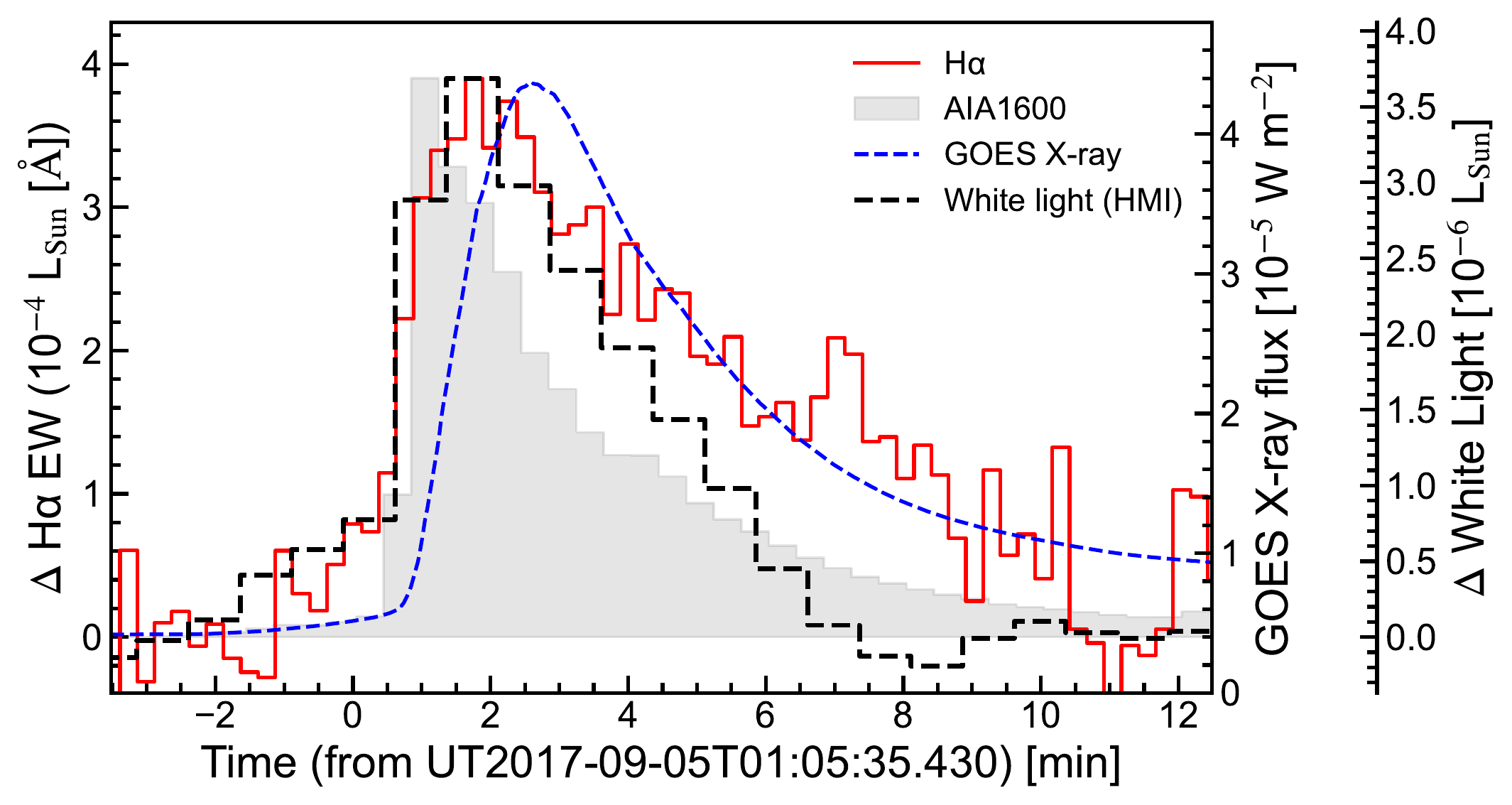}
\epsscale{0.46}
\plotone{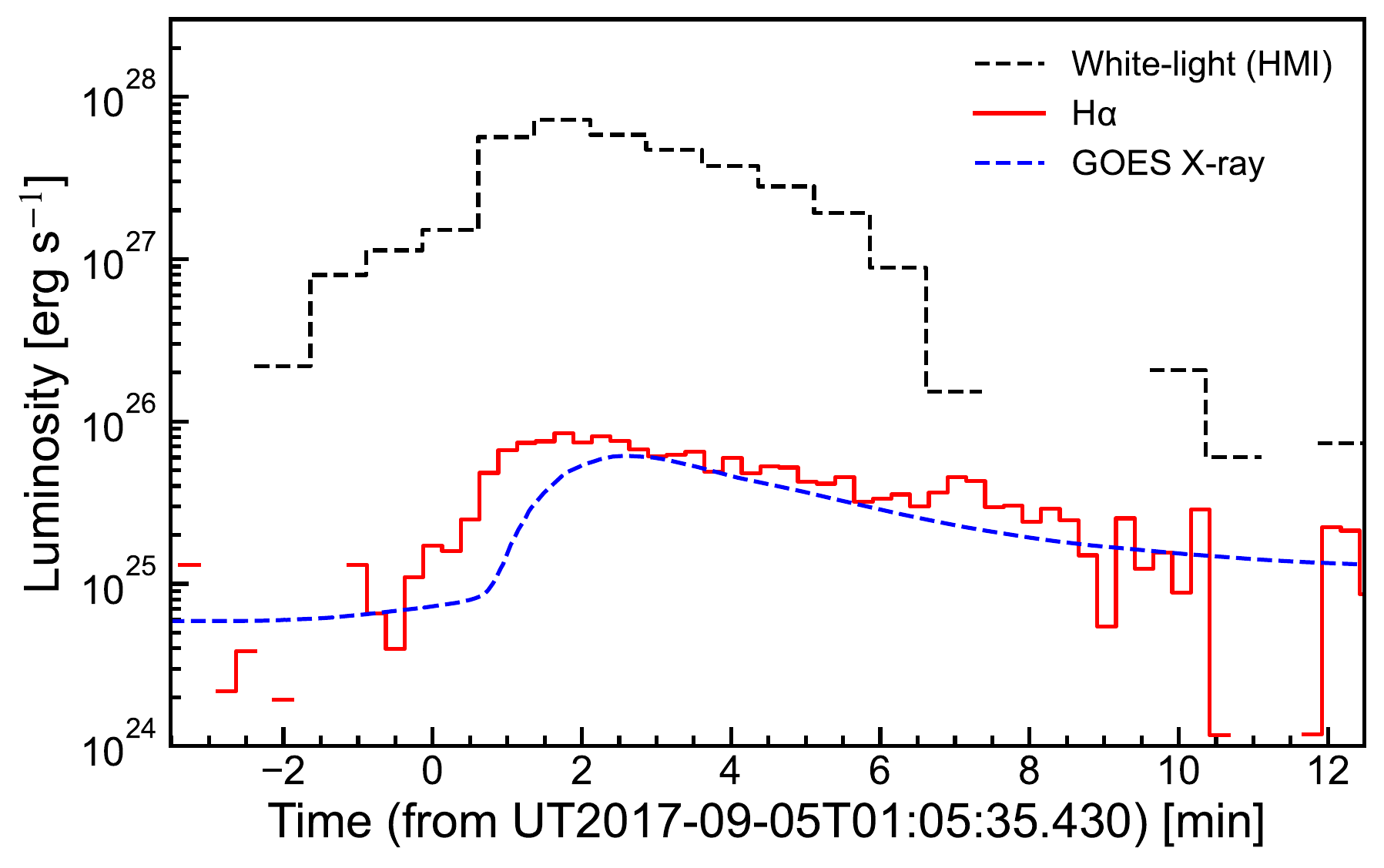}
\caption{(left) The Sun-as-a-star light curves of the solar flare in H$\alpha$ equivalent width (EW), white light, GOES X-ray, and AIA 1600 {\AA} in each unit. Note that the AIA 1600 {\AA} light curve is normalized by the peak of the H$\alpha$ flare. (right) The Sun-as-a-star luminosity variation of the solar flare in H$\alpha$, white light, and GOES X-ray in logarithmic scale. 
}
\label{fig:6}
\end{figure}

\begin{figure}
\epsscale{0.5}
\plotone{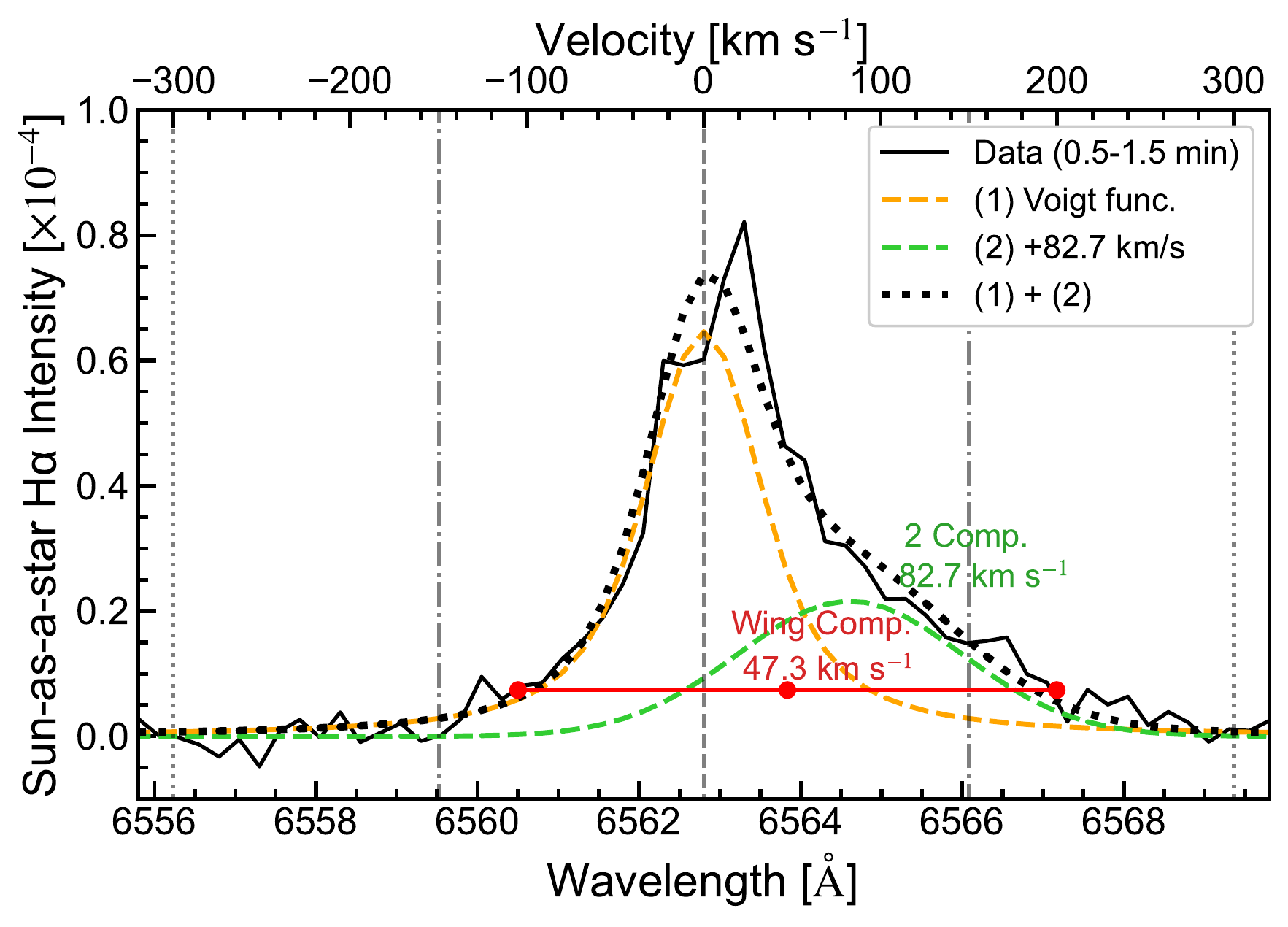}
\caption{An example of fitting result for a Sun-as-a-star spectrum (0.5-1.5  minutes) with 2 component fitting (orange and green lines) and single wing component fitting (red markers). 
A black dotted line presents the sum of the orange and green lines. In this case, the former method estimates the redshift velocity as 82.7 km s$^{-1}$, and the latter estimates it as 47.3 km s$^{-1}$.
}
\label{fig:7}
\end{figure}

\begin{figure}
\epsscale{1}
\plottwo{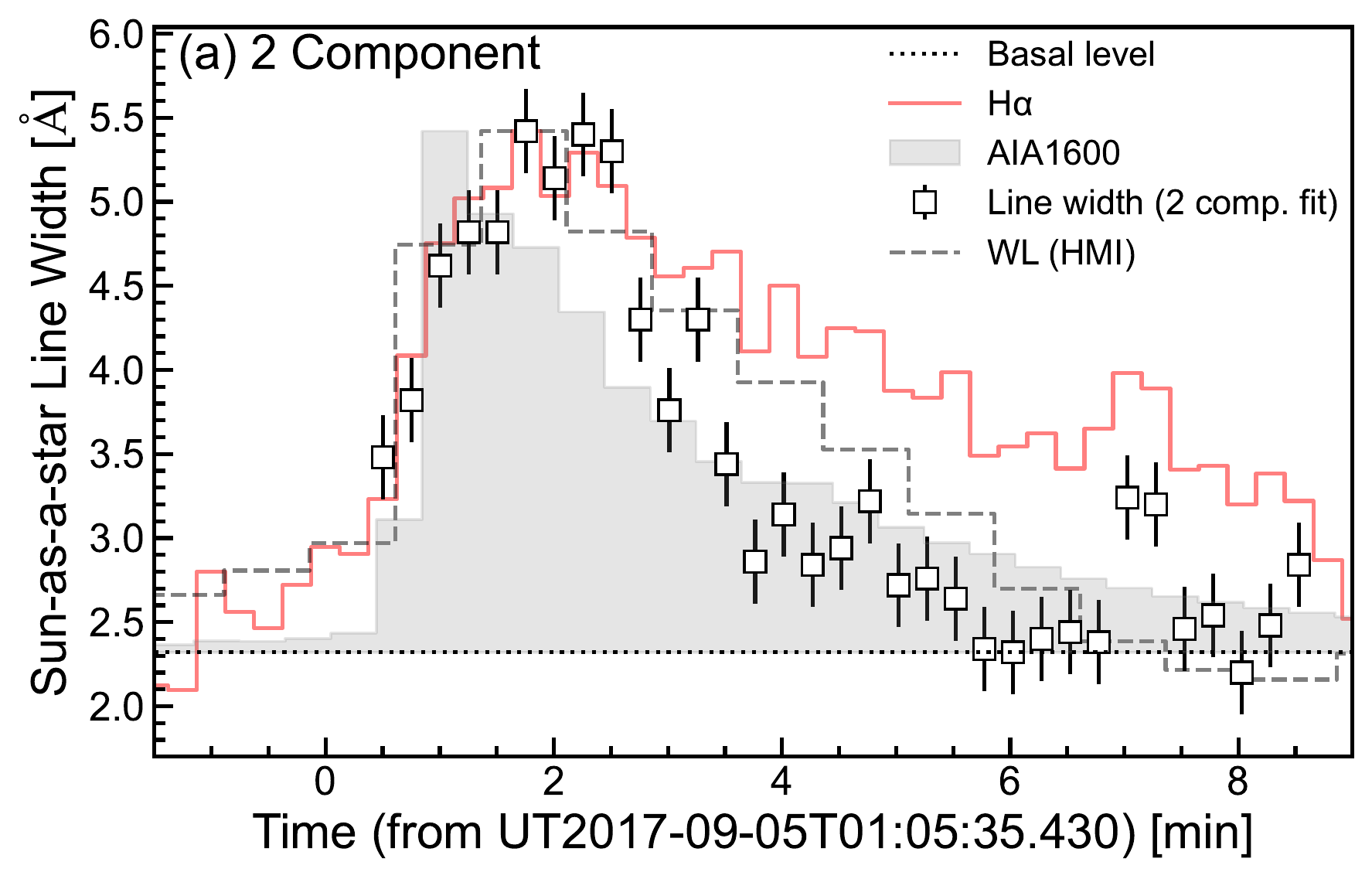}{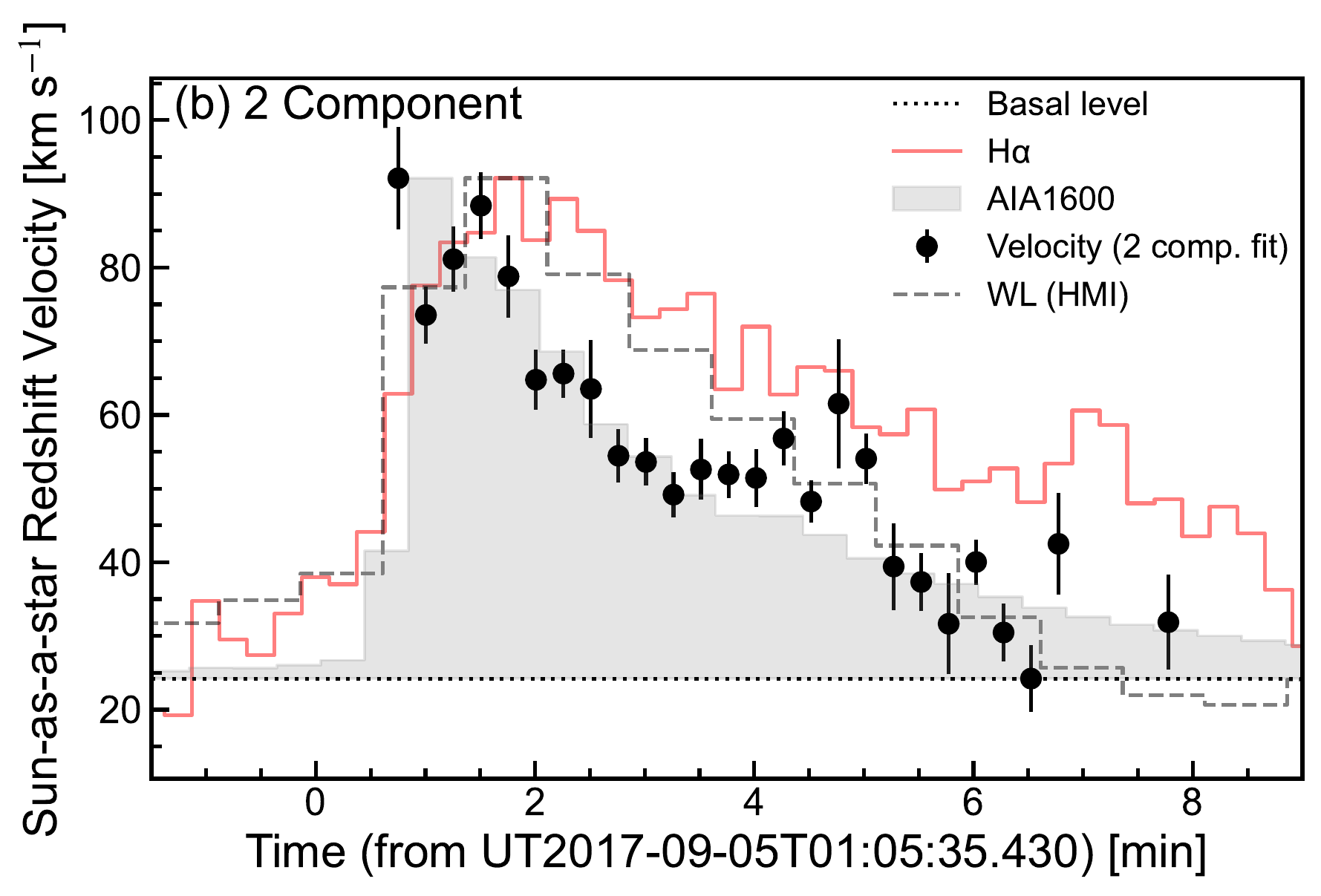}
\plottwo{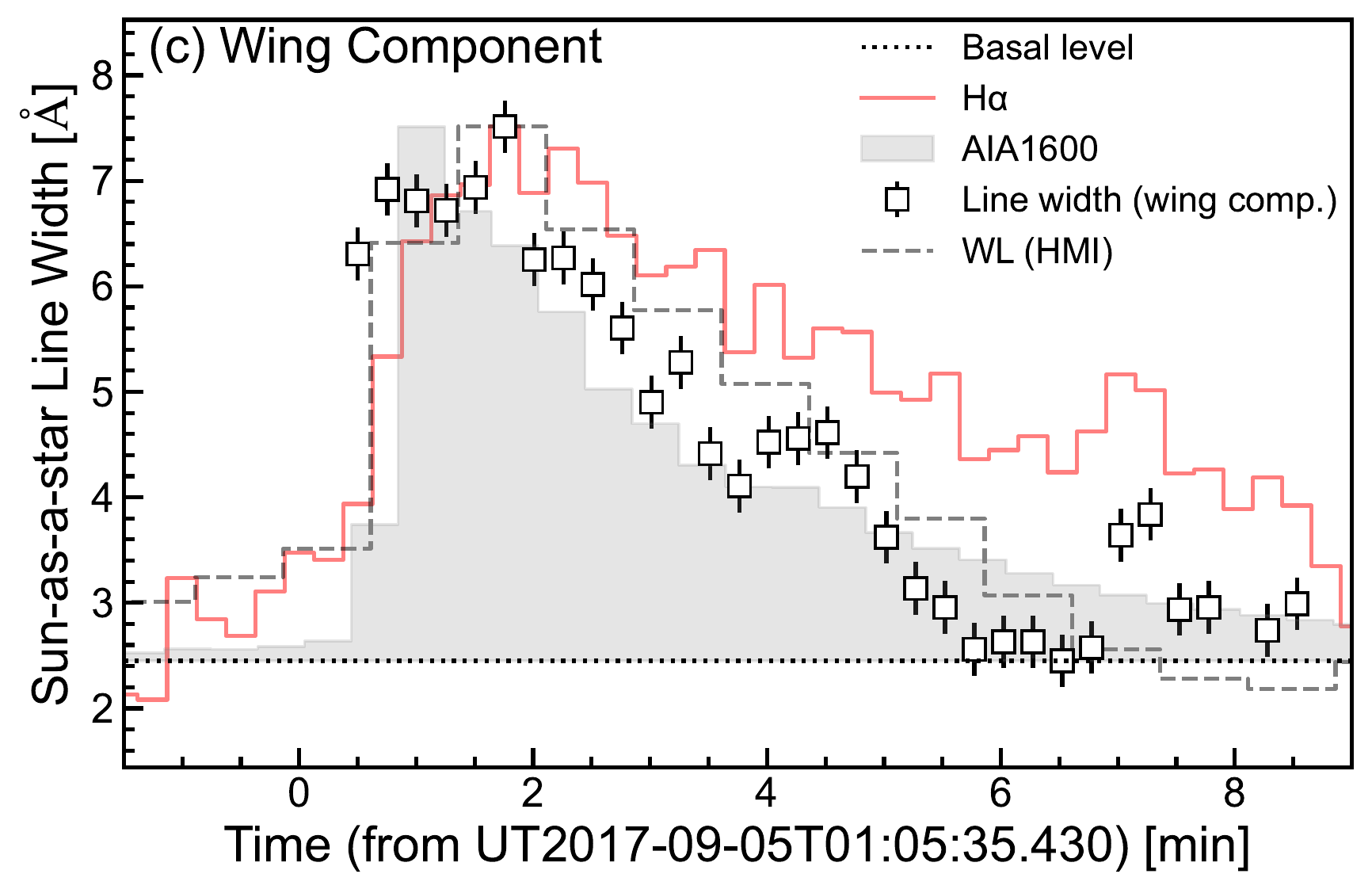}{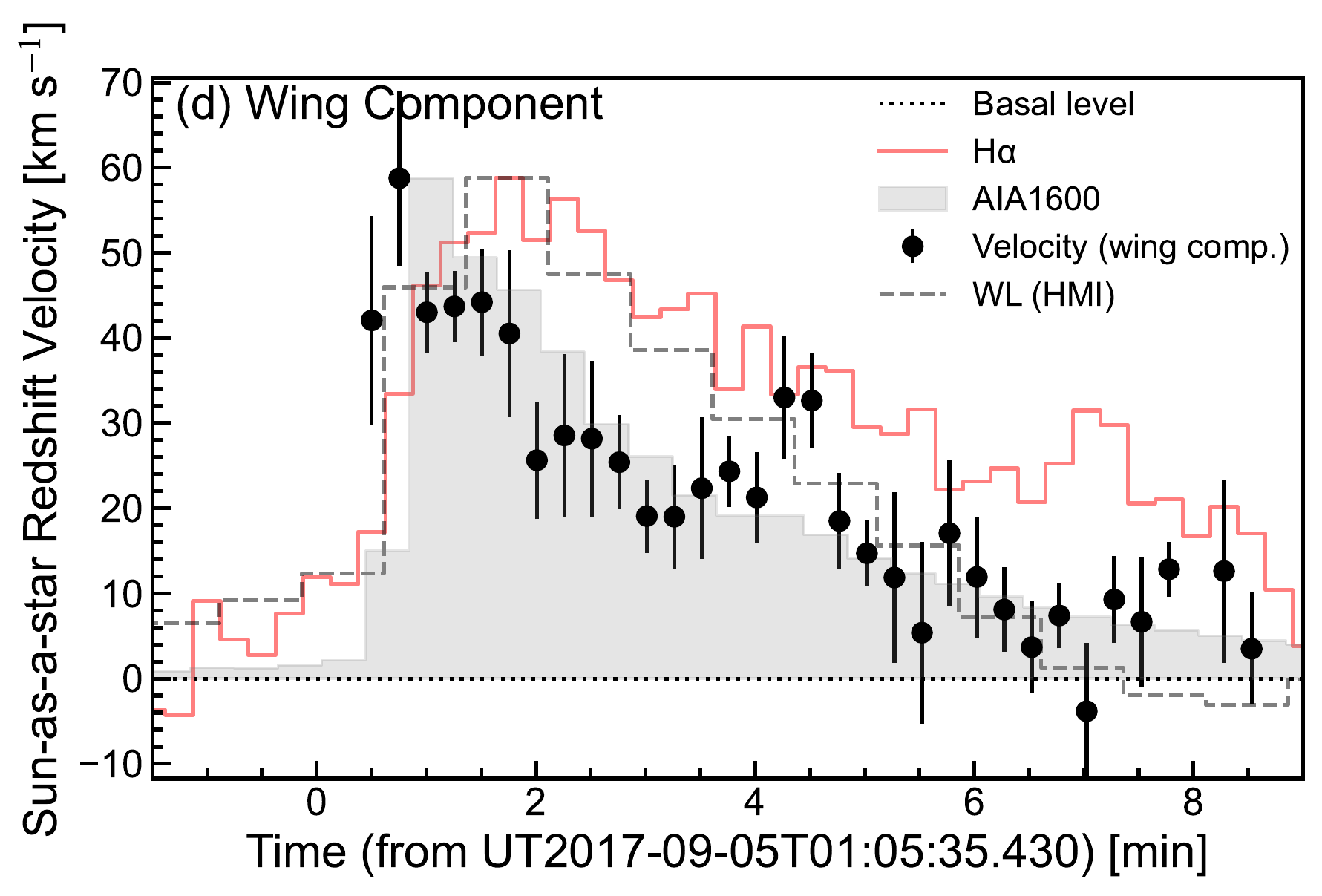}
\epsscale{0.5}
\plotone{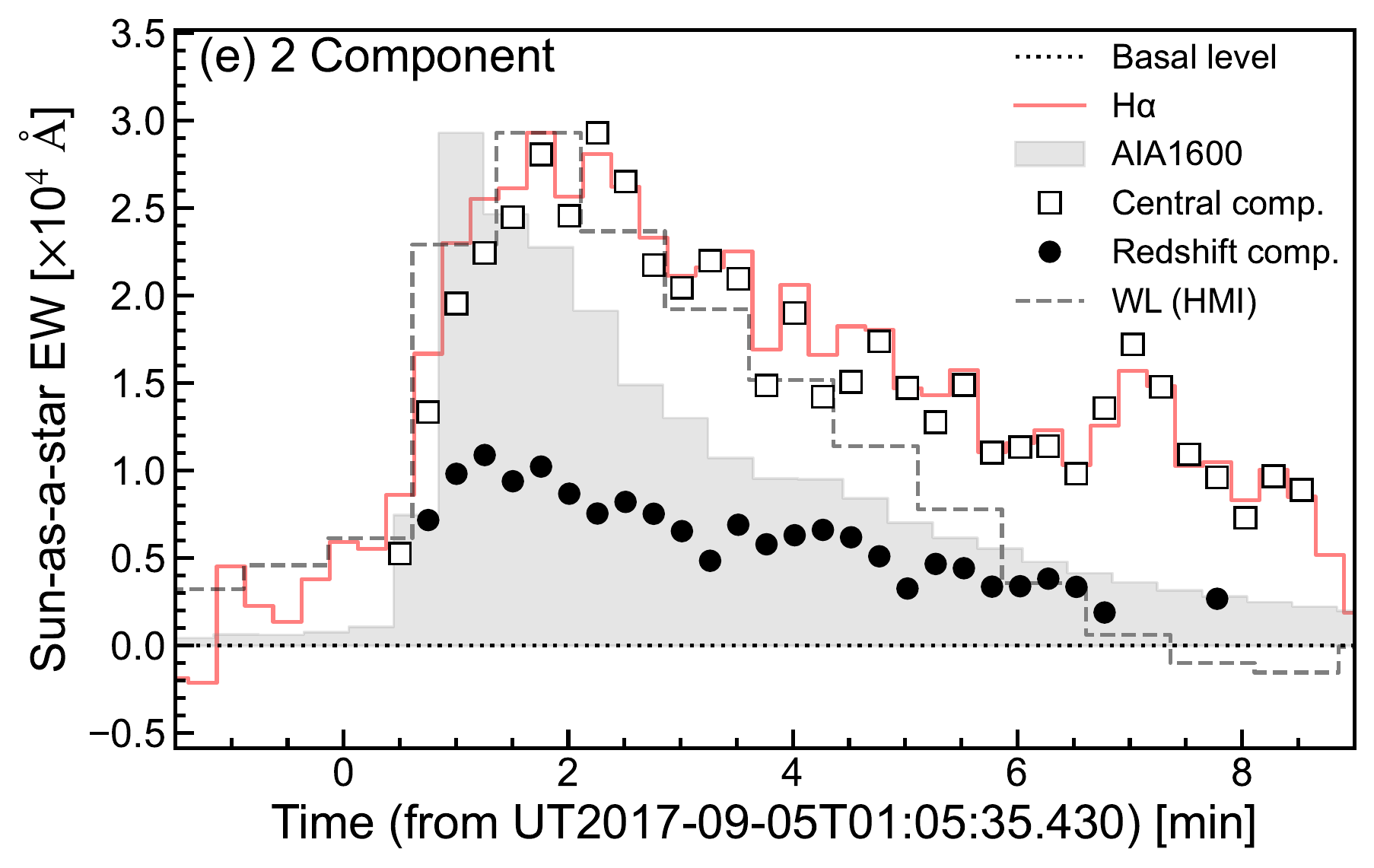}
\caption{Temporal evolution of the line width, redshift velocity, and equivalent width of redshift component of Sun-as-a-star H$\alpha$ spectra. The normalized light curve in H$\alpha$ and white light are plotted with the red solid line and black dashed line, respectively, as references. (a) The line width estimated with 2 component fitting. The error bar shows just a spectral resolution. (b) The redshift velocity estimated with 2 component fitting. The error bar shows fitting errors. (c) The line width estimated with single wing component fitting. (d) The redshift velocity estimated with single wing component fitting. (e) The EW of the central component (open square) and redshift component (filled circles) estimated with 2 component fitting.
Note that the basal levels of (a), (b), and (c) are defined as the minimum values in each panel ($t$ $<$ 7 minutes), which are consistent with the line width of original spectra of H$\alpha$ line $\sim$2.5 {\AA}.
}
\label{fig:8}
\end{figure}

Left panel of Figure \ref{fig:6} shows Sun-as-a-star light curves of the solar flare in H$\alpha$, white light, GOES X-ray, and AIA 1600 {\AA}.
The flaring amplitude in Sun-as-a-star H$\alpha$ EW is $\sim4\times10^{-4}$ {\AA}.
This is approximately the same order of the impulsive C-class flare event, i.e. $\sim3\times10^{-4}$ {\AA}, in our previous study \citep[see, Supplementary Figure 9 in][]{Namekata2020Sci}.
The duration of the H$\alpha$ flare was approximately 10 minutes (FWHM duration is 4.7 minutes), and that of white-light flare is approximately 7 minutes (FWHM duration is 3.0 minutes). 
The AIA 1600 {\AA} flare peak came first, followed by the H$\alpha$ and white light about a minute later, and then the GOES X-rays about a minute later.
Rising phase of the AIA 1600 {\AA} and GOES X-ray light curves have a relation suggested by the \textit{Neupert} effect \textcolor{black}{(see Figure \ref{fig:13} in Appendix \ref{sec:app-1})}, so the AIA 1600 {\AA} light curve can be regarded as a proxy of time profile of energy release of the flare.

The right panel of Figure \ref{fig:6} shows the light curve of flare luminosity in H$\alpha$, white light, and GOES X-ray (1--8 {\AA}).
The flare luminosity of H$\alpha$ (8.4$\times 10^{25}$ erg s$^{-1}$) and GOES X-ray (6.1$\times 10^{25}$ erg s$^{-1}$) have almost the same order ($\sim$10$^{26}$ erg s$^{-1}$).
The flare luminosity of white light (7.2$\times 10^{27}$ erg s$^{-1}$) is an about two orders of magnitude larger than those of H$\alpha$ and GOES X-ray (1--8 {\AA}).
The integrated flare radiated energy are estimated as $1.6\times$10$^{30}$ erg in white light and $2.8\times$ 10$^{28}$ erg in H$\alpha$ (i.e. 1.8 \% of the white-light emission).
Since few studies have measured the energy of white light and H$\alpha$ in solar flares, such values are worth to be compared with those of stellar flares \citep[e.g.,][]{2015ApJ...809...79O,2019A&A...622A.210G}.
For example, the ratio of H$\alpha$ to white-light energy $\sim$1.8 \% are comparable to those of stellar superflares on solar-type stars \citep[0.85$\sim$1.5 \%;][]{Namekata2020Sci,2022arXiv220109416N}.

\subsection{Redshift velocity and line width in Sun-as-a-star spectra}\label{sec:3-2}

Figure \ref{fig:7} shows the example of a result of the spectral fitting for a Sun-as-a-star flare spectrum with the two fitting methods introduced in Section \ref{sec:2-4} and Figure \ref{fig:4}.
In the case of Figure \ref{fig:7}, the 2 component fitting methods estimates the redshift velocity as 82.7 km s$^{-1}$, and the single wing component fitting methods estimates as 47 km s$^{-1}$.
Figure \ref{fig:8} shows the time evolution of each parameter for Sun-as-a-star H$\alpha$ spectrum in comparison with H$\alpha$, white-light, and AIA 1600 {\AA} light curve.
The line width is ranging from 2.5 {\AA} to 7.5 {\AA}, and the redshift velocity is from 0 km s$^{-1}$ to 95 km s$^{-1}$ during the solar flare (see, Section \ref{sec:2-4}).
The two fitting methods result in different absolute values, but the qualitative properties are roughly similar.
In particular, the Sun-as-a-star spectra shows the following properties:
\begin{itemize}
\item The redshift velocity peaks out and decays more rapidly than the H$\alpha$ EW, and the time evolution is similar to near-ultraviolet luminosity (1600 {\AA}), rather than H$\alpha$ and white light (Figure \ref{fig:8}(b,d)).
\item The H$\alpha$ line width also peaks out and decays more rapidly than the H$\alpha$ EW, and the time evolution is nearly consistent to white-light luminosity, rather than H$\alpha$ and near ultraviolet (1600 {\AA}) (Figure \ref{fig:8}(a,c)).
\item The time evolution of EW of redshifted component is similar to that of EW of the central component (Figure \ref{fig:8}(e)), and the ratio of redshifted component to central component is 14$\sim$54 \% during the flare.
\end{itemize}

\subsection{Redshift velocity and line width in spatially-resolved spectra}\label{sec:3-3}

\begin{figure}
\epsscale{0.5}
\plotone{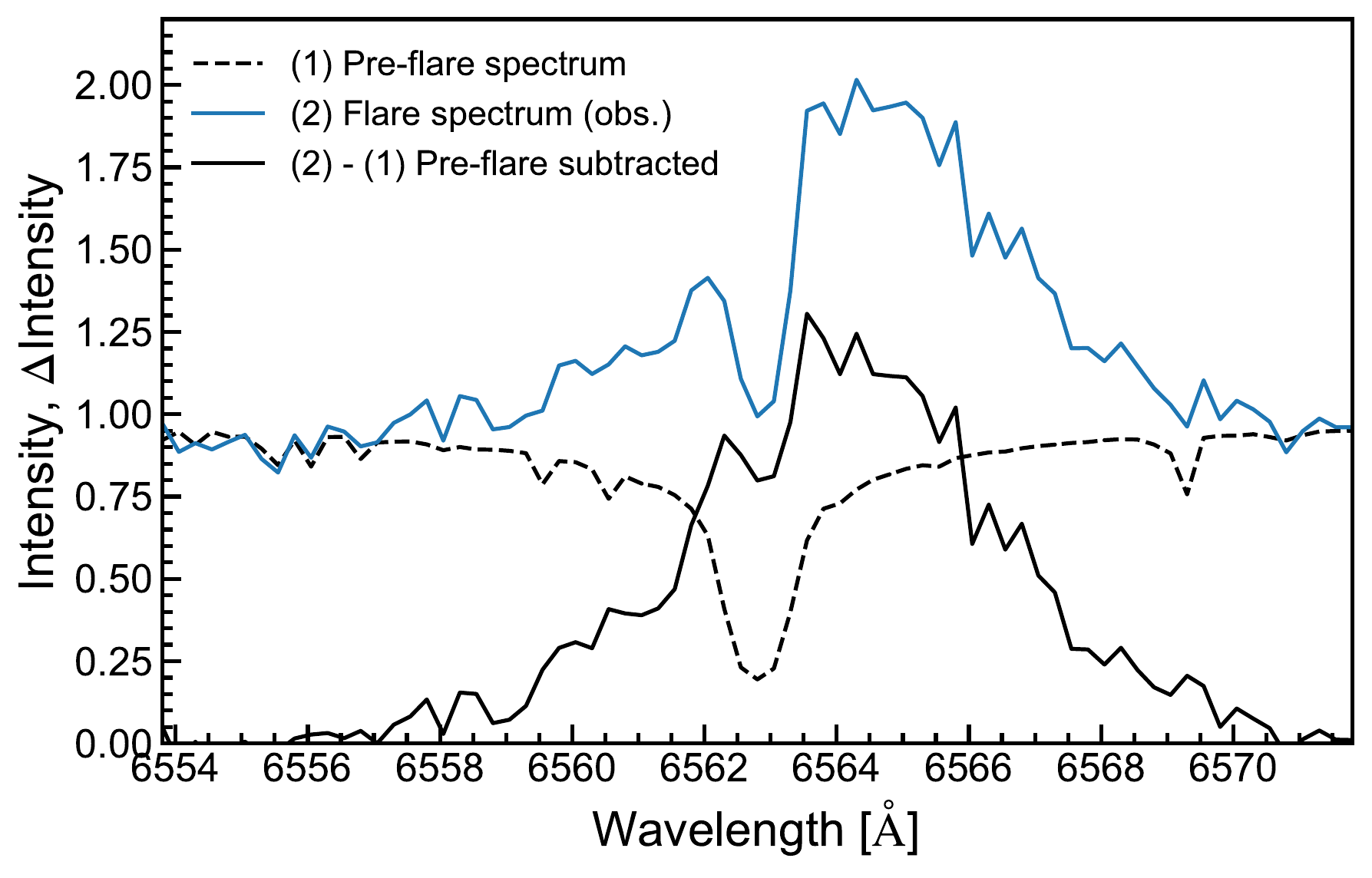}
\caption{
\color{black} An example of the pre-flare subtraction of the H$\alpha$ spectrum at a given flare kernel ($\times$ point in Figure \ref{fig:9}) and a given time (01:06:36 UT). The black dashed line is the background H$\alpha$ spectrum. The blue solid line is the H$\alpha$ spectrum of the flare, and the black solid line is the pre-flare-subtracted H$\alpha$ spectrum. 
}
\label{fig:8.5}
\end{figure}

\begin{figure}
\epsscale{1}
\plotone{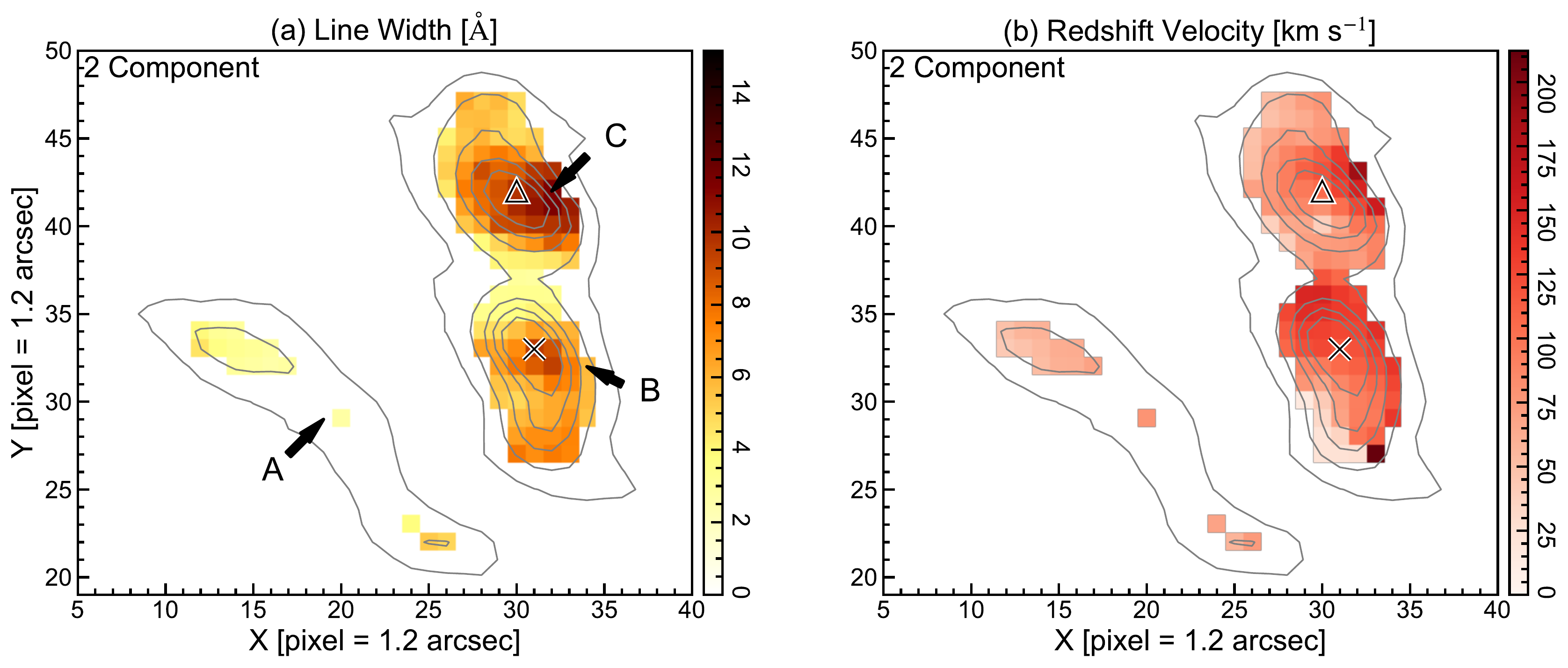}
\plotone{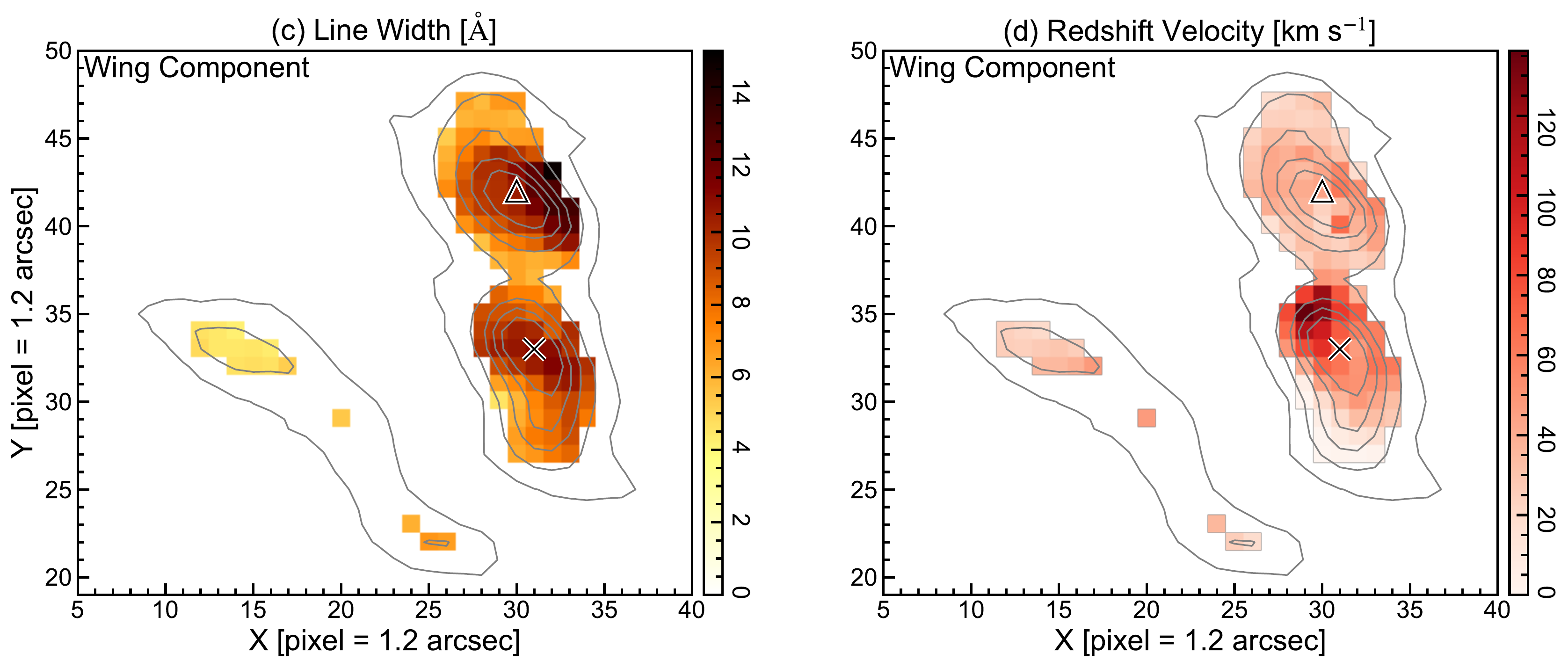}
\caption{
An example of spatial distribution of the line width and redshift velocity of H$\alpha$ lines in the flare ribbon at 01:06:36 UT (t=1.0  minute) on 2017 September 5.
The contour shows the EW level (5, 15, 30, 45, 60 \% from the outside) against the maximum-EW pixel at this time.
(a) The line width estimated with 2 component fitting. (b) The redshift velocity estimated with 2 component fitting.  (c) The line width estimated with single wing component fitting. (d) The redshift velocity estimated with single wing component fitting.
For the triangles and crosses, time evolution are indicated in Figure \ref{fig:10}. 
In panel (a), the three flare ribbons A, B, and C are labeled with the arrows (see Figure \ref{fig:2} and \ref{fig:3})
}
\label{fig:9}
\end{figure}

\begin{figure}
\epsscale{1}
\plottwo{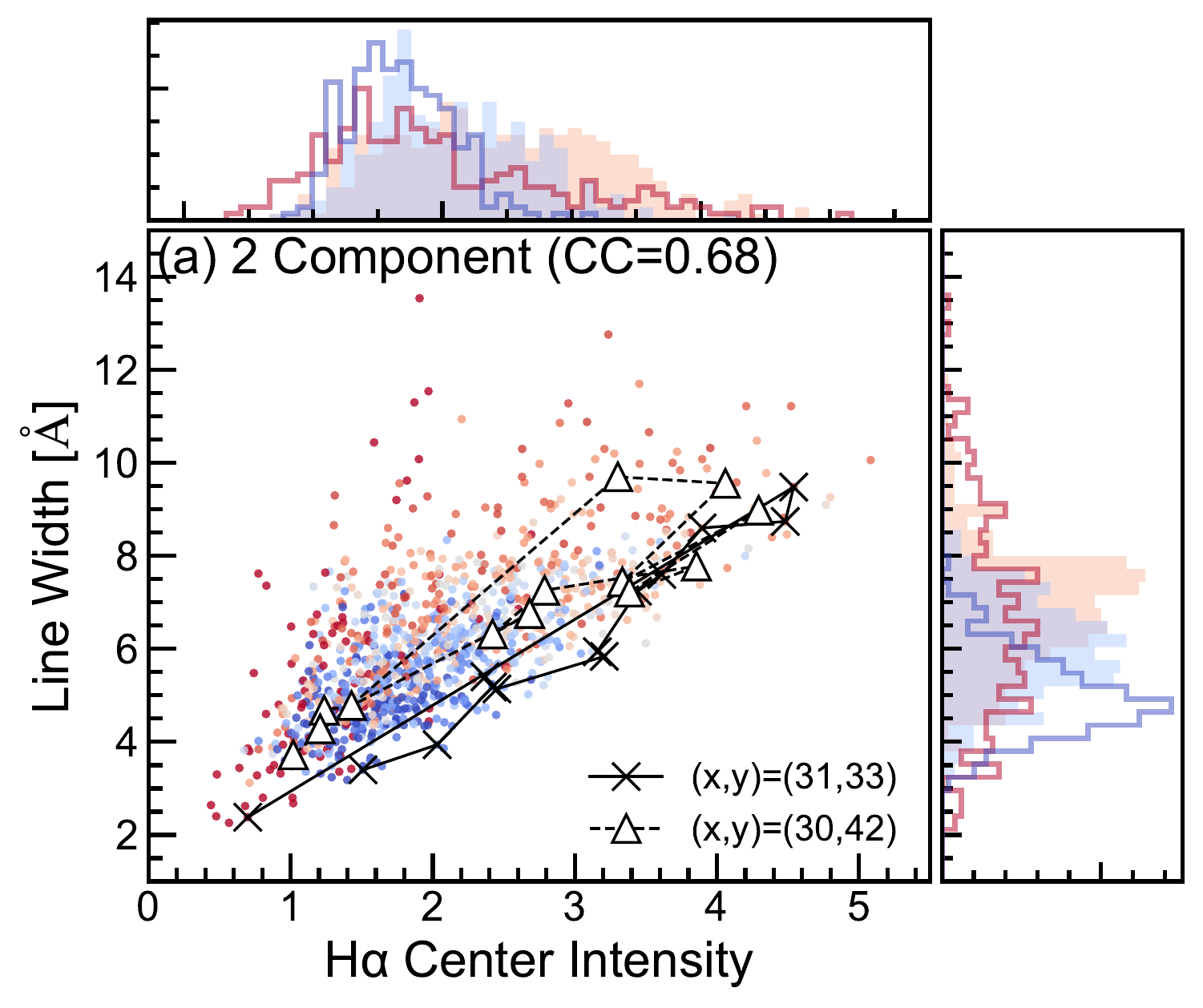}{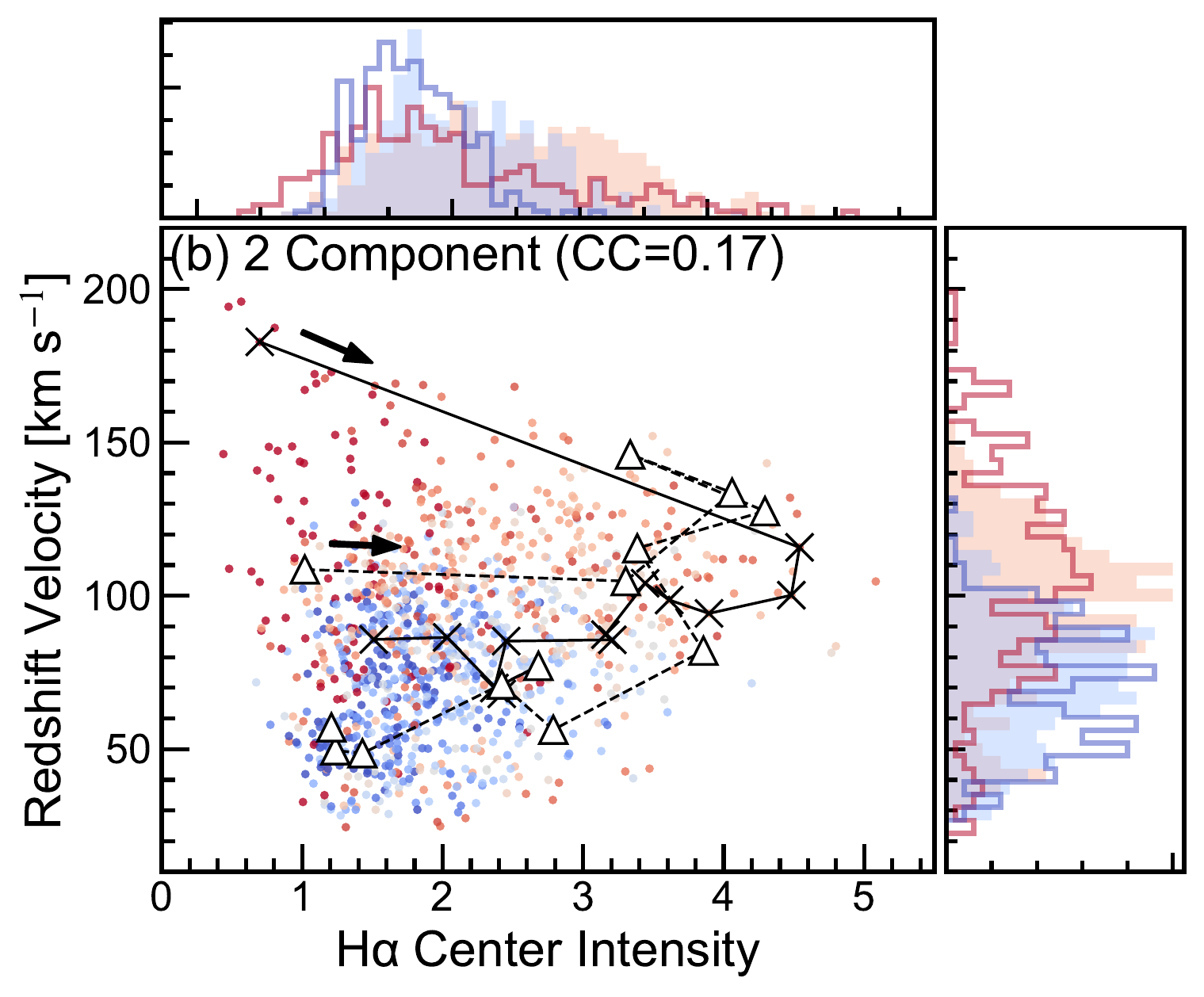}
\plottwo{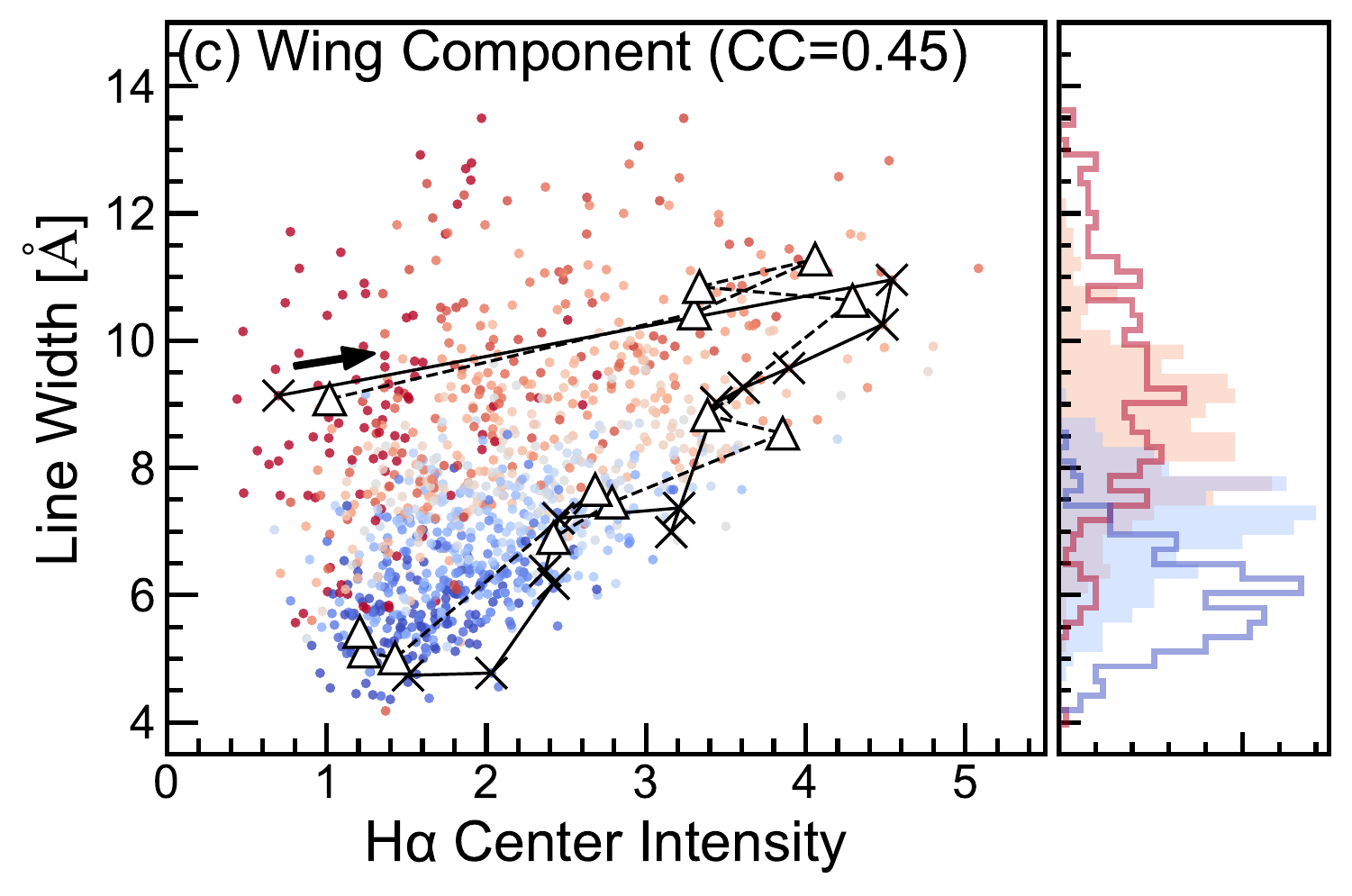}{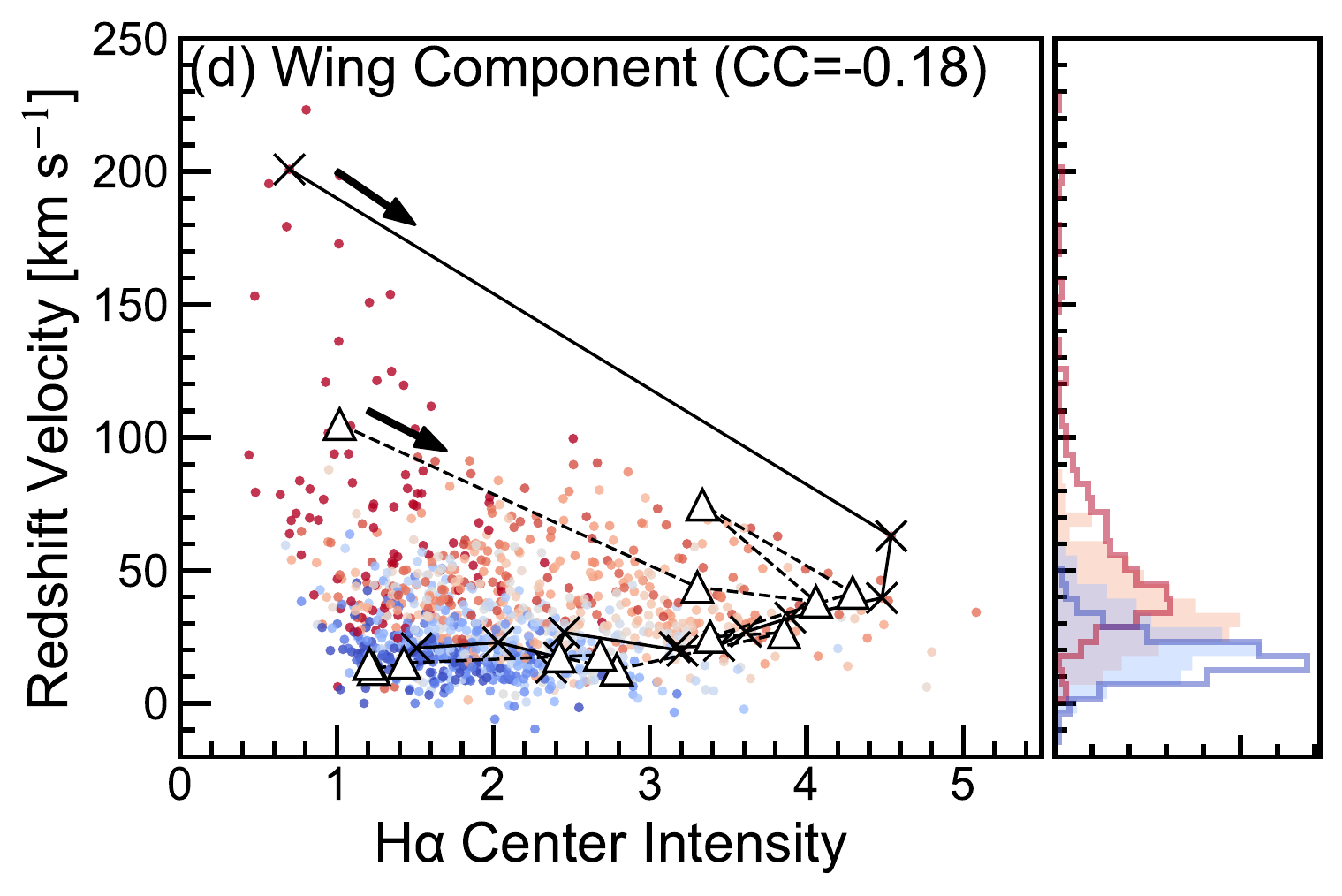}
\epsscale{0.5}
\plotone{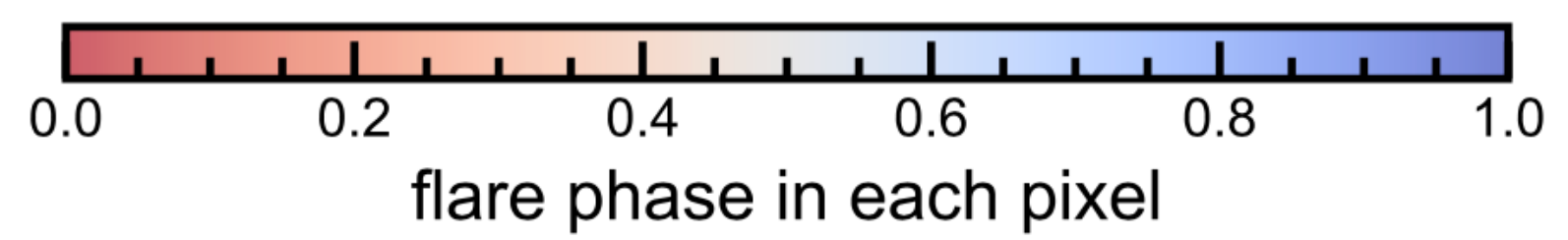}
\caption{.
Time evolution of line width and redshift velocity as a function of H$\alpha$ center intensity for each pixel  and time. 
(a) The line width estimated with 2 component fitting. (b) The redshift velocity estimated with 2 component fitting.  (c) The line width estimated with single wing component fitting. (d) The redshift velocity estimated with single wing component fitting.
The color represents the flaring phase ``for each pixel" as indicated in the color bar. 
The triangles and crosses are examples of time evolution of pixels indicated in Figure \ref{fig:9}. 
Correlation coefficient (CC) is described in each panel. 
}
\label{fig:10}
\end{figure}

\textcolor{black}{Figure \ref{fig:8.5} shows an example of the pre-flare subtraction of the spatially-resolved H$\alpha$ spectrum at a given flare kernel and a given time.
As explained in Section \ref{sec:2-4}, the pre-flare subtraction can apparently enhance the central component of the pre-flare-subtracted H$\alpha$ spectrum of flares (a black solid line) because the background (a black dashed line) is absorption, but would not significantly affect the evaluation of shifted and/or broadened component (see Section \ref{sec:2-4}).
We also analyzed the pre-flare-subtracted spectrum for spectral fitting in this spatially-resolved case. }
Figure \ref{fig:9} shows the spatial distribution of the line width (a,c) and redshift velocity (b,d) of H$\alpha$ lines in the flare ribbon obtained by applying the two methods for each pixel.
We fitted the H$\alpha$ spectrum for each pixel and each time with the different two methods, and derived the temporal evolution of the spatial distribution.
Figure \ref{fig:10} shows the line width (a,c)  and redshift velocity (b,d) of H$\alpha$ lines as a function of H$\alpha$ line center intensity.
The estimated line width is ranging from $\sim$2.5 {\AA} to $\sim$14 {\AA}, and the redshift velocity is from 0 km s$^{-1}$ to $\sim$230 km s$^{-1}$.
The cross and triangle symbols indicate the time evolution of the pixels indicated in Figure \ref{fig:9} with the same symbols ((x,y) = (31,33) and (30,42), respectively). Also, the color of each dot corresponds to the phase of the flare in each pixel (see the color bar).
Histograms for earlier and later phases are shown on the right and on the top with red and blue lines.

We found that the line width has a correlation with the intensity of H$\alpha$ (correlation coefficient of 0.68 and 0.45 for the two methods), but the redshift velocities have almost no correlation with the intensity of H$\alpha$ (correlation coefficient of 0.17 and $-$0.18 for the two methods).
\cite{2012PASJ...64...20A} also showed that red asymmetry (RA; Equation \ref{eq:ra}) is independent of intensity, although they did not derive the velocity itself.
On the other hand, if we look at each pixel individually, we can see that there is an evolutionary track through the flare phases.
For example, if we look at the triangle and cross symbols, the evolutionary track is that line width and redshift velocity increase first, then intensity increases later, and finally they gradually decay.
This is apparent in Figure \ref{fig:10}(b,c,d), but not apparent in Figure \ref{fig:10}(a).
Especially for redshift velocity, \cite{2012PASJ...64...20A} and \cite{1984SoPh...93..105I} reported the same property, i.e., redshift velocity peaks out and decays more rapidly than intensity.

\subsection{Comparison between Sun-as-a-star spectrum and spatially-resolved spectra}\label{sec:3-4}

\begin{figure}
\epsscale{0.5}
\plotone{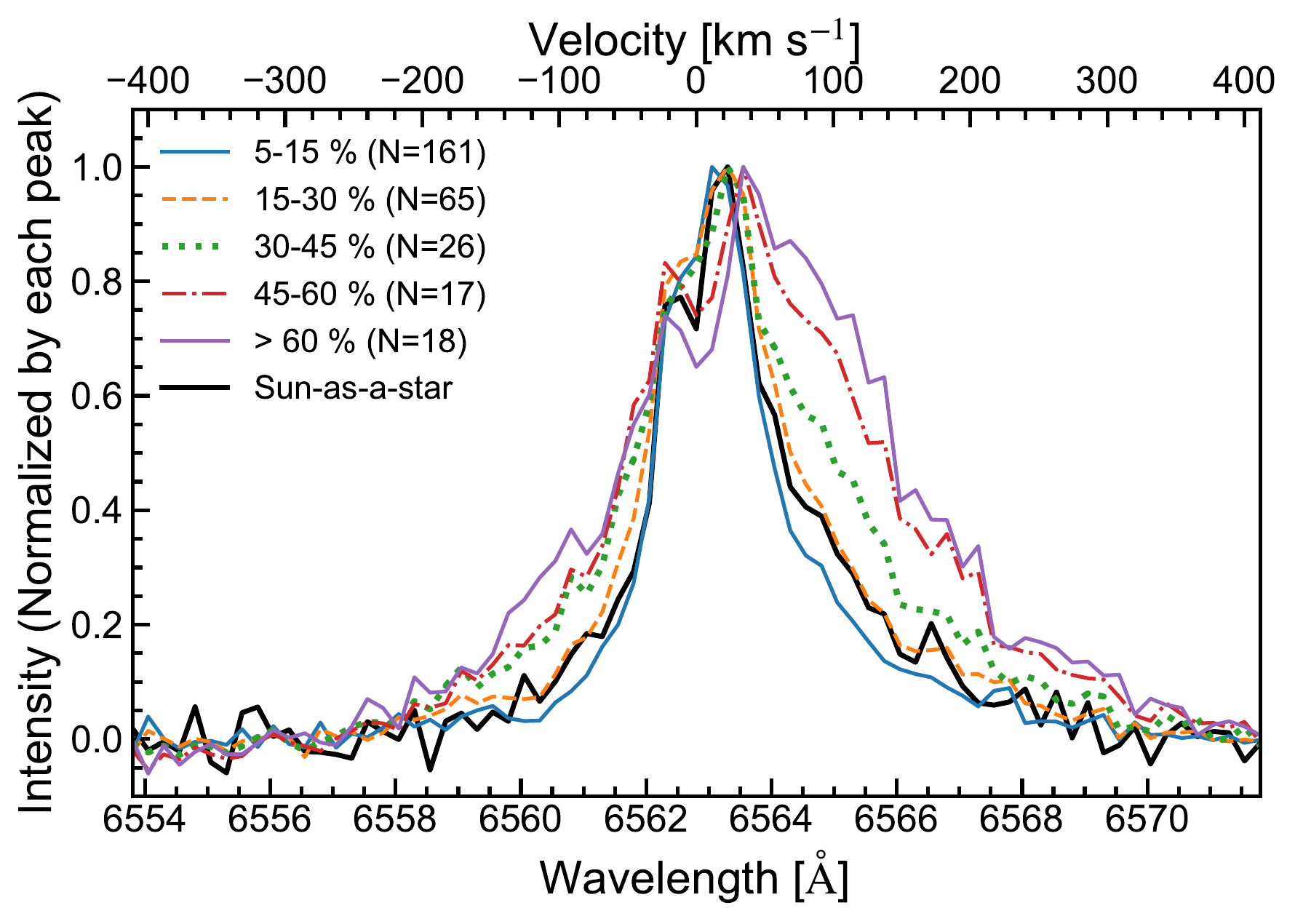}
\plotone{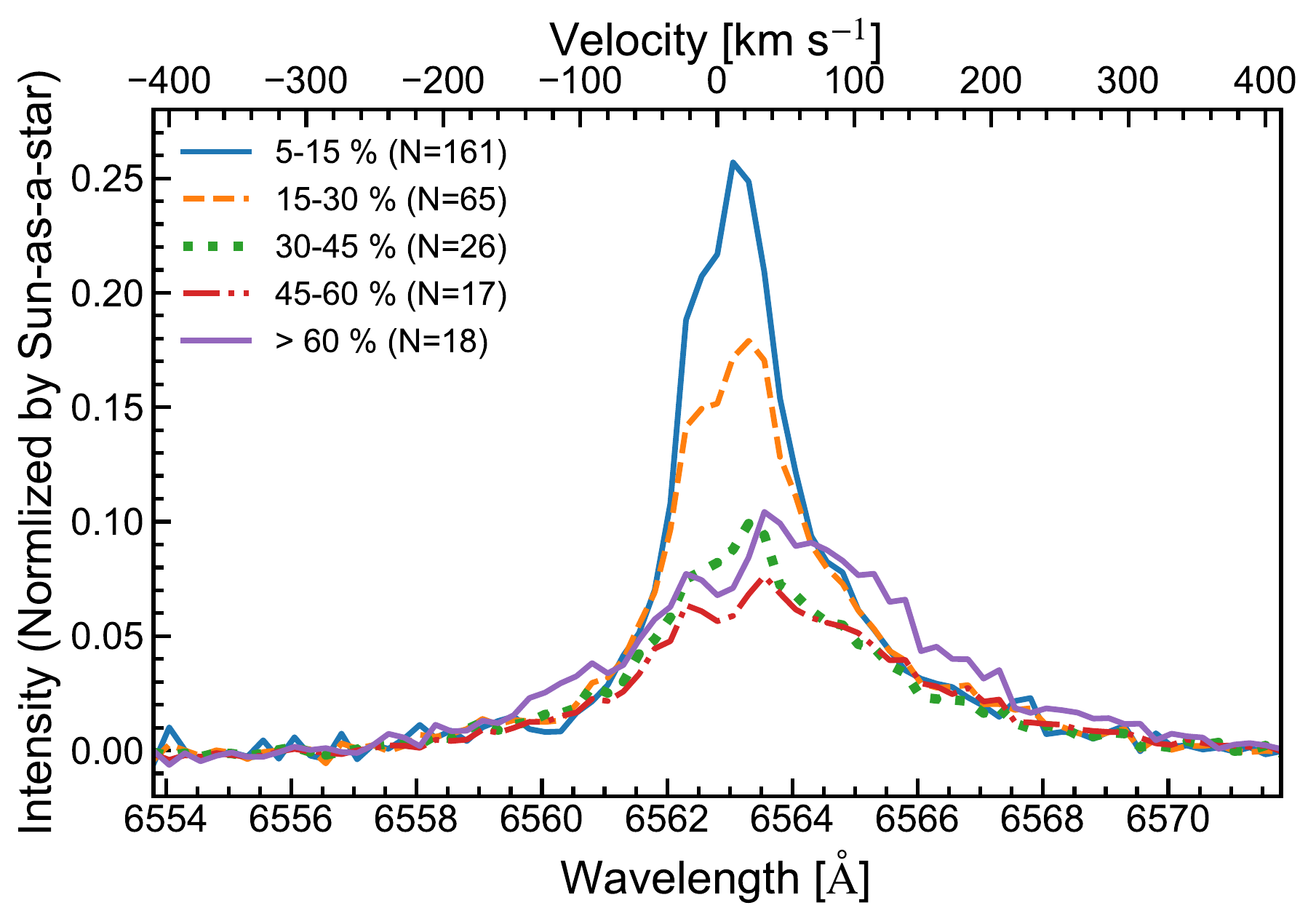}
\caption{Spatially-integrated spectra inside a different H$\alpha$ intensity (EW) levels (each colored spectrum) at 01:06:36 UT (t=1.0  minute) on 2017 September 5 in comparison with Sun-as-a-star spectra (a black line).
The integrated spectrum inside the region of 5-15 \%, 15-30 \%, 30-45 \%, 45-60 \%, and $>$60 \% levels of the maximum-EW pixel at the time is plotted  (see the contour in Figure \ref{fig:10}). (left) A comparison of spectra normalized by each spectral peak. (right) A comparison of the spatially-integrated spectra normalized by the peak of the Sun-as-a-star spectrum.
}
\label{fig:11}
\end{figure}

\begin{figure}
\epsscale{1}
\plottwo{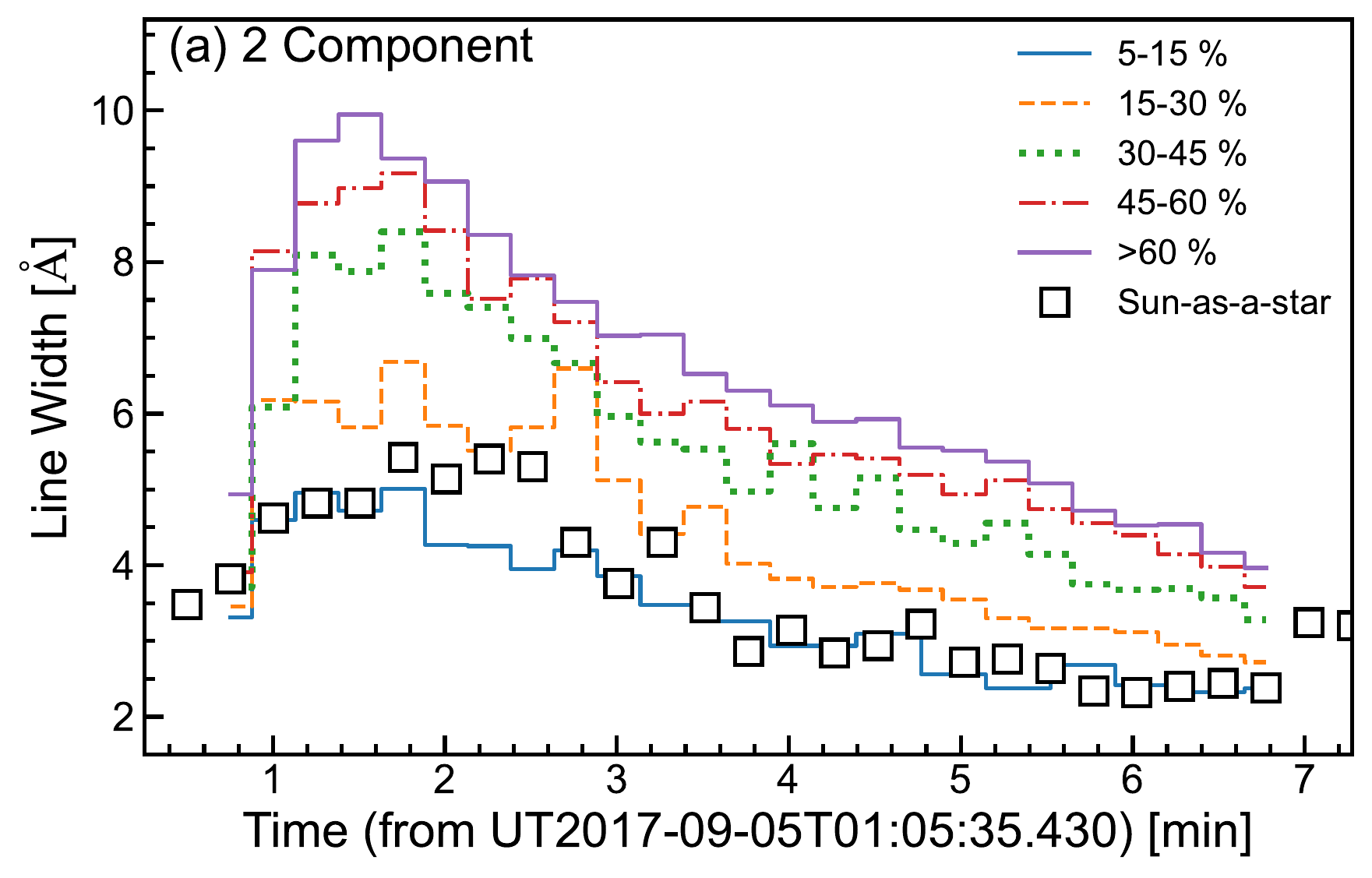}{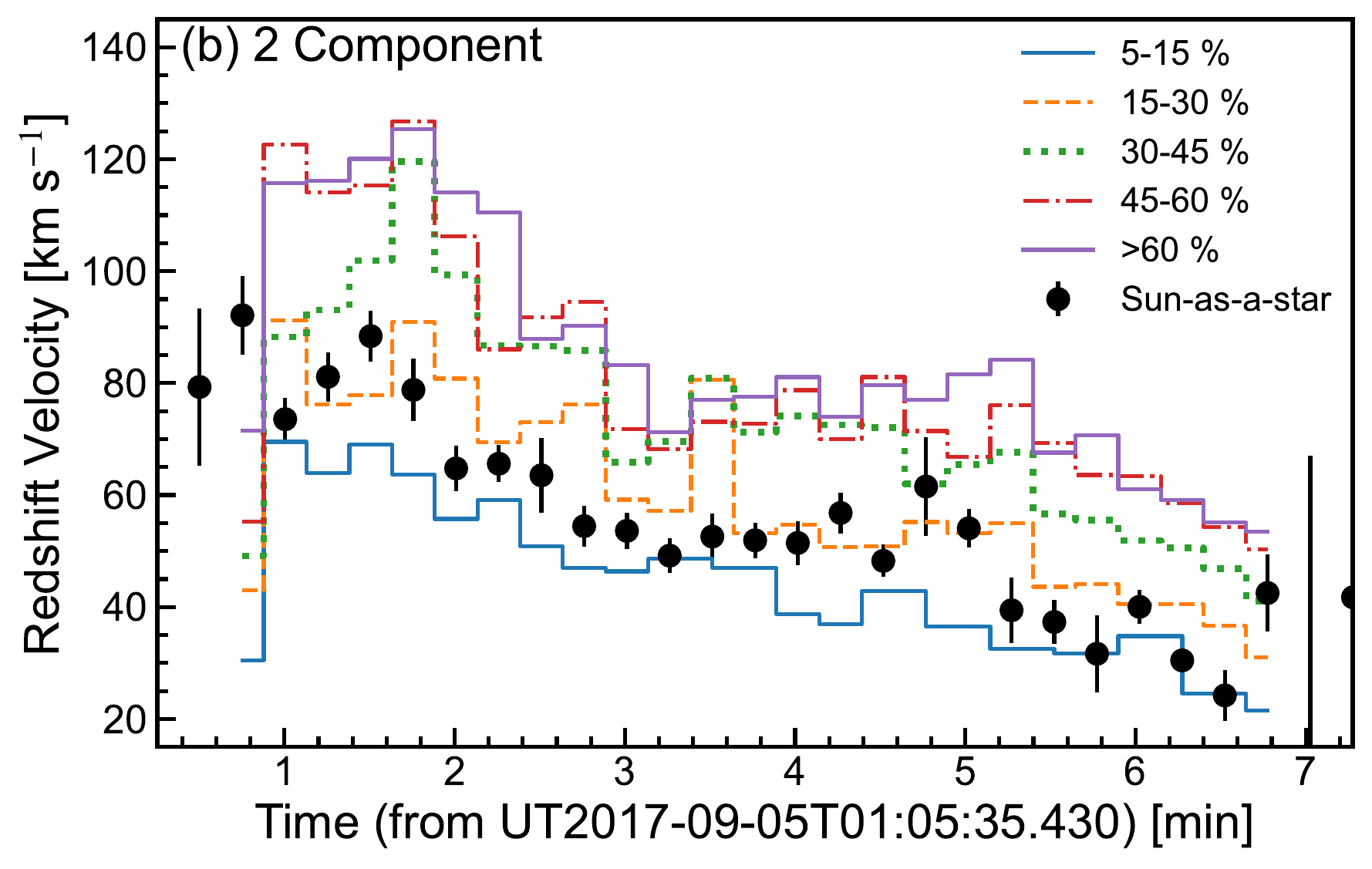}
\plottwo{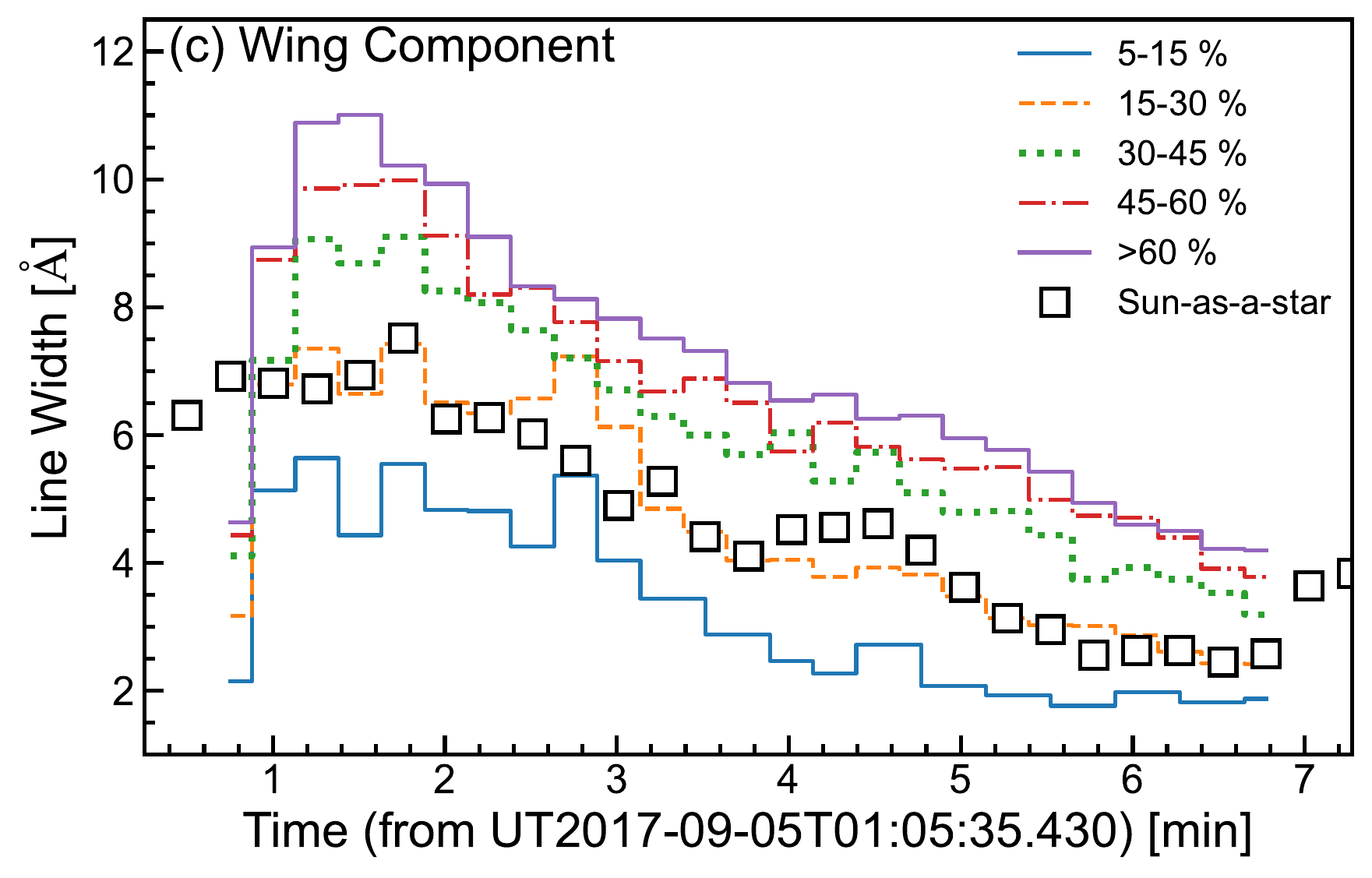}{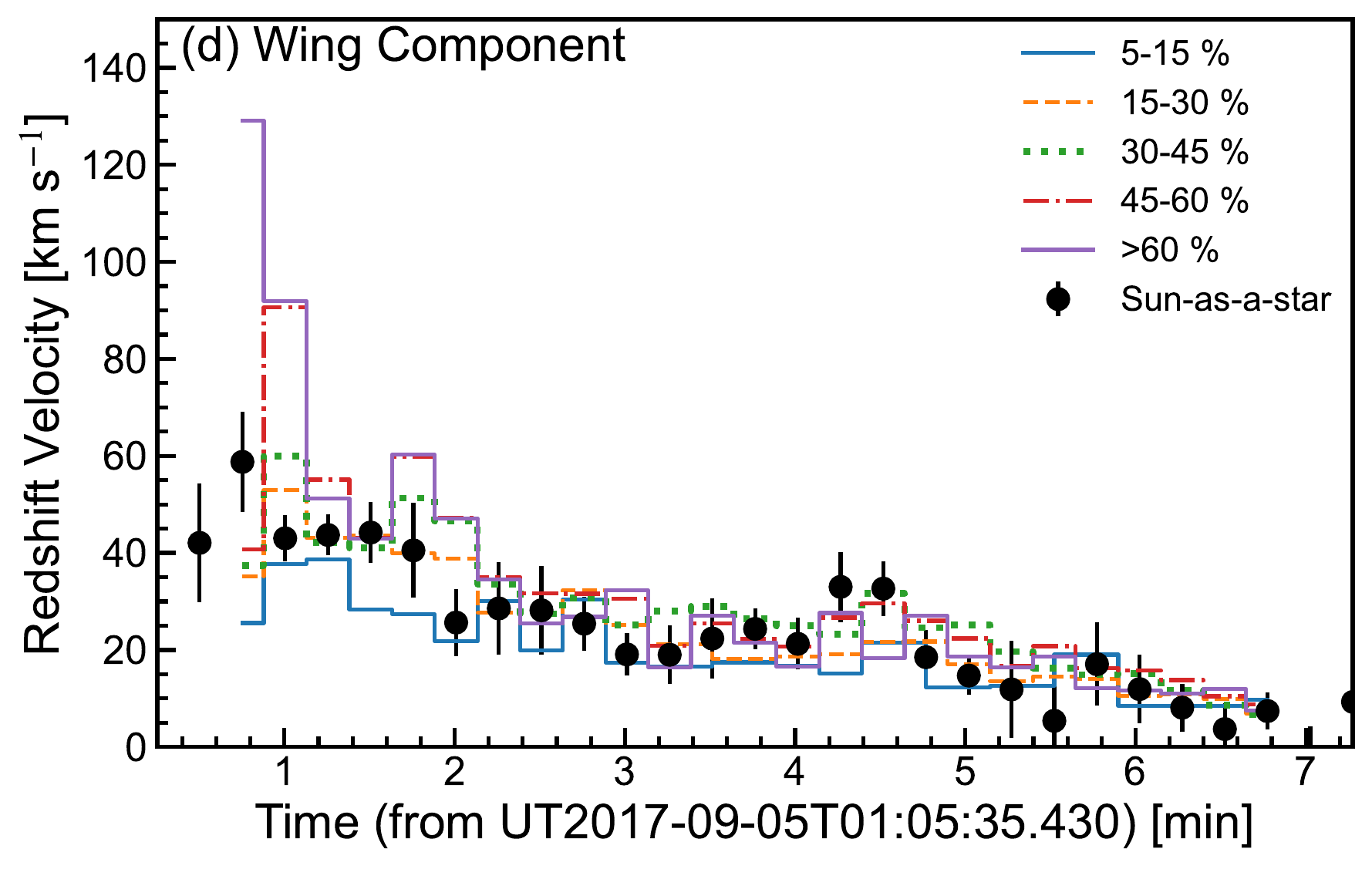}
\caption{Comparison of line width and redshift velocity between the Sun-as-a-star spectra (black symbols) and local spatially-integrated spectra (colored lines; see the caption in Figure \ref{fig:12}). 
(a) The line width estimated with 2 component fitting. (b) The redshift velocity estimated with 2 component fitting.  (c) The line width estimated with single wing component fitting. (d) The redshift velocity estimated with single wing component fitting.
}
\label{fig:12}
\end{figure}

Here we compare line width and redshift calculated from Sun-as-a-star (Section \ref{sec:3-2}) spectra and spatially-resolved spectra (Section \ref{sec:3-3}).
Qualitatively, the variations of both spectra show very similar evolutionary track: line width and redshift velocity peak out and decay more rapidly than EW.
On the other hand, the spatially resolved case has significantly wide range of values, while the Sun-as-a-star case has relatively smaller values than the upper limit of the spatially-resolved case.
This would be because the Sun-as-a-star spectrum is seen as a superposition as discussed in Section \ref{sec:4-1} in more detail.

Then, which regions of the flare represent the Sun-as-a-star spectrum?
The left panel of Figure \ref{fig:11} shows the comparison between Sun-as-a-star spectrum (black line) and integrated spectra for a different H$\alpha$ intensity levels at a given time.
This shows that the Sun-as-a-star spectrum is quite far from the spectrum from the region of high intensity level ($>$45 \% of the maximum intensity, N$_{\rm pixel}$=35 pixels), which shows strong redshift and line broadening.
On the other hand, the Sun-as-a-star spectrum seems to correspond fairly well with the spectrum from the region where the intensity level is weak, especially 15-30 \% intensity level (N$_{\rm pixel}$=65 pixels).
The right panel of Figure \ref{fig:11} is the same as the left panel, but it is normalized by the peak of Sun-as-a-star spectrum.
This shows that where the intensity level is weak, the area of the region (number of pixels) is large, so the contribution to the Sun-as-a-star spectrum is high.
Therefore, as a result of the superposition, the Sun-as-a-star spectrum becomes close to the shape of the spectrum from the region with the weak intensity level (in this case, 15-30 \%).

Figure \ref{fig:12} shows the time evolution of line width and redshift velocity for each intensity levels. 
The comparison shows that the Sun-as-a-star spectrum well corresponds to the spectrum from 15-30 \% (or, 5-15 \%) intensity level.
It is also found that the line width of the Sun-as-a-star spectrum is about a factor of 2 (2-5 \AA) smaller than the strong-intensity region of the flare ribbon (e.g. $>$ 60 \%). 
Moreover, in the case of the two component fitting, the redshift velocity of the Sun-as-a-star spectrum is about 20-40 km s$^{-1}$ slower than that of the core region of the flare ribbon (e.g. $>$ 60 \%). 
Note that the redshift velocities obtained by single wing component fitting correspond well to the Sun-as-a-star spectra in all regions except the brightest core at the flare onset.

\section{Discussion}\label{sec:4}

\subsection{Formation of Sun-as-a-star H$\alpha$ flare spectra}\label{sec:4-1}

Section \ref{sec:3-4} shows that the region with weak intensity level (here we call ``halo") is more significantly contributing to Sun-as-a-star spectrum than the region with strong intensity level (here we call ``core").
It is true that the core of the flare ribbon is bright, but its contribution to the Sun-as-a-star radiation is not necessarily large.
This indicates that the red asymmetry and line width that can be observed in stars do not represent the bright ``core" component but ``halo" component, and underestimates the ``core" value by a factor of $\sim$2.
Our first message to stellar flare observation is that we need to keep in mind the above point when we interpret stellar flare spectra and compare with numerical model \citep[e.g.,][]{2017ApJ...837..125K,2020PASJ...72...68N,2022ApJ...928..180W}.
Note that only for the redshift velocity calculated with the single-wing-component fitting gives good correspondence between Sun-as-a-star values and spatially-resolved values (see Figure \ref{fig:12}(d)). 
This may indicate that the single-wing-component fitting can better estimates the evolutions of flare ribbons in stellar flare observations, although it still may miss the highest velocity component at the onset of flares.

\subsection{Time evolution of the redshift velocity of Sun-as-a-star H$\alpha$ flare spectra}\label{sec:4-2}

In Section \ref{sec:3-2}, we show the evolutionary track of the redshift velocity of the Sun-as-a-star H$\alpha$ spectra, i.e., redshift velocity peaks out and decays more rapidly than H$\alpha$ intensity or EW.
Section \ref{sec:3-3} and \ref{sec:3-4} shows the same feature seen even in the spatially-resolved spectrum, \textcolor{black}{which was also previously reported not only in H$\alpha$ \citep[e.g.,][]{1984SoPh...93..105I} but also in other chromospheric and transition-region lines \citep[e.g.,][]{2020ApJ...904...95Z}.}
These mean that Sun-as-a-star H$\alpha$ spectra of solar flares well reflect the feature of the spatially-resolved H$\alpha$ spectrum, even though it is formed as a result of superposition of different positions as discussed in Section \ref{sec:4-1}. 
According to previous studies, the fact that the increase in redshift velocity precedes the increase in intensity is thought to represent the properties of chromospheric condensation process, where thickness of the chromospheric condensation increase with the rapid decrease of its velocity \citep{1981SoPh...73..269L,1982SoPh...81..281S,1984SoPh...93..105I,2022arXiv220113349K}.
Our study suggests that even the Sun-as-a-star spectra retain the characteristics of chromospheric condensation.
This predicts that we can investigate occurrences of chromospheric condensation even in stellar flares from the time-resolved H$\alpha$ spectroscopic observations.

The redshift velocity shows similar changes to Sun-as-a-star light curve of the near-ultraviolet rays at 1600 {\AA}, rather than Sun-as-a-star light curve of white-light flare.
The AIA 1600 {\AA} has a sensitivity to the transition region radiation (C IV lines) and its time evolution shows very impulsive peak during the impulsive phase, which is represented by the time derivative of the GOES X-ray \citep{2009ApJ...694L..74N,2013ApJ...774...14Q,2018ApJ...863..124D}.
Our study suggests that the time evolution of redshift velocity can be a good proxy for heating in the upper chromosphere and transition region (or possibly non-thermal heating) in the Sun-as-a-star view.
Although hard X-rays, an indicator of non-thermal heating, is not observable in stellar flares, the near-ultraviolet wavelength can be observed in stellar flares by \textcolor{black}{the Ultraviolet Optical Telescope onboard} the Swift spacecraft \citep{2010ApJ...721..785O} and \textcolor{black}{the Optical/UV Monitor Telescope onboard} XMM-Newton \citep[cf.,][]{2005A&A...431..679M}, so we propose that these features could be tested in stellar flares in the future.

Recently, there are reports on red asymmetry on stellar flares \citep{1993A&A...274..245H,2019A&A...623A..49V,2020arXiv201200786K,2022ApJ...928..180W}, but its origin has not been well understood partly because of the lack of time-resolved spectroscopic observations or multi-wavelength observations are limited. 
Our study can provide a basis for interpreting the observed red asymmetry on stellar flares.
If the time variation of redshift of stellar flares is the same as that of this result, we can conclude that it comes from flare ribbon radiation, while if it is not, we can suggest the possibility of other sources of redshift radiation, such as flare loop or prominence activation.
For example, \cite{2022ApJ...928..180W} reported that redshifted velocity of H$\alpha$ line asymmetry during a superflare on an M-type star showed more rapid decay than H$\alpha$ EW. 
This observation indicates that the observed red asymmetry of the superflare may originate from chromospheric condensation region.
\textcolor{black}{Here, we need to be a bit careful when applying our result on the Sun to other stars whose background H$\alpha$ line shows emission (active M-type stars) since we performed an analysis of pre-flare subtraction of spectra in this study.
If the background is emission, pre-flare-subtraction is thought to subtract the H$\alpha$ line center component of the flare more than necessary.
In this case, although the discussion of the qualitative nature of line asymmetry/broadening (velocity and line width) is applicable to stars (see Section \ref{sec:2-4}), the ratio of line center intensity to red asymmetry intensity is expected to be different from those obtained in this solar study.}

\subsection{Time evolution of line width of Sun-as-a-star H$\alpha$ flare spectra}\label{sec:4-3}

Sun-as-a-star H$\alpha$ spectra shows that the line width decays more rapidly than EW (Section \ref{sec:3-2}), and the spatially-resolved spectra show the same feature (Section \ref{sec:3-3} and \ref{sec:3-4}).
This also supports the suggestion that Sun-as-a-star H$\alpha$ spectra of solar flares well inform the feature of the spatially-resolved H$\alpha$ spectrum even after the superposition. 
Similar time evolution has been reported on M-dwarf flares \citep[e.g.,][]{2020PASJ...72...68N,2022ApJ...928..180W}, and this study suggests that the chromospheres on the M-dwarf flares could be heating in the same way as the M4.2-class solar flare analyzed here.


It is also found that the line width peaks out and decays in the similar way to the Sun-as-a-star light curve of white light flare (or near-ultraviolet flare), rather than H$\alpha$ EW.
In the case of solar flares, line broadening is thought to be caused mainly by the statistical Stark effect \citep{2017ApJ...837..125K,2022arXiv220113349K}, and the line width is sensitive to electron density $n_e$ in chromospheric condensation ($\propto n_e^{2/3}$).
Also, if we assume that the white-light emission is radiated from the same chromospheric condensation via hydrogen free-bound or free-free process, its radiation intensity reflects the chromospheric density ($\propto n_e^{2}$).
Therefore, the nearly similar time evolution of white-light flare and line width may be understood as both are related to the density of chromospheric condensation.
\textcolor{black}{Note, however, that the origin of white-light emission is still controversial even in solar physics, further studies are necessary for the physical interpretation.} 



\section{Summary and Future Work}\label{sec:5}

In this paper, we present the Sun-as-a-star analysis of H$\alpha$ line of the M4.2-class solar flare showing dominant emission from flare ribbons (i.e. chromospheric radiation) by using the data of SMART/SDDI, with the aim of future applications to the interpretations of stellar flares.
This Sun-as-a-star analysis can provide a standard example for interpreting observed H$\alpha$ spectra of stellar flares and distinguishing flare ribbon radiations and other phenomena.
The Sun-as-a-star light curves of white-light flares (from heated chromosphere/photosphere) and \textcolor{black}{near-ultraviolet} flares (1600 {\AA}, mostly from heated transition region), which can be also observable in stellar flares, were also obtained as a proxy of impulsive radiations.
The results and implications to stellar flare studies are summarized as follows: 
\begin{itemize}
\item[(i)] Sun-as-a-star H$\alpha$ spectra of the solar flare show a red asymmetry and line broadening. This is well known when flares are spatially resolved, but has not been reported in Sun-as-a-star view. The redshift velocity ranges up to $\sim$95 km s$^{-1}$, and the line broadening ranges from $\sim$2.5 {\AA} to $\sim$7.5 {\AA} (Section \ref{sec:3-2}).
\item[(ii)] Sun-as-a-star H$\alpha$ spectra of the solar flare is consistent with the spectra from weak-intensity flare ribbons like a ``halo", rather than the spectra from strong-intensity flare ribbons ``core". As a results, the line width and redshift velocity in Sun-as-a-star spectra are smaller by a few factors (typically, $\sim$2) than those of the core component, depending on the fitting methods. We need to care this result when we interpret stellar flare observations (Section \ref{sec:3-4} and \ref{sec:4-1}). 
\item[(iii)] The EW of H$\alpha$ emission of the flare is an order of 10$^{-4}$ {\AA} in Sun-as-a-star spectrum. This is similar to those of C-class flares reported by \cite{Namekata2020Sci}. The total radiated energies of H$\alpha$ and white-light flare is 2.8$\times$10$^{28}$ erg and 1.6$\times$10$^{30}$ erg, respectively. The ratio of H$\alpha$ to white-light energy $\sim$1.8 \% are similar to those of stellar superflares on solar-type stars \citep{Namekata2020Sci,2022arXiv220109416N}. See Section \ref{sec:3-1}. 
\item[(iv)] The redshift velocity peaks out and decays more rapidly than the H$\alpha$ EW. The preceding of redshift velocity to intensity can be seen in spatially-resolved data and has been interpreted as a nature of chromospheric condensation \citep{1981SoPh...73..269L,1982SoPh...81..281S,1984SoPh...93..105I,2022arXiv220113349K}. This means that, even after the superposition, the nature of chromospheric condensation is visible in the Sun-as-a-star spectrum, and possibly in stellar flare spectrum. The features are actually observed in M-dwarf flares \textcolor{black}{(Namizaki et al. 2022, submitted)}. Also, the time variation (rise and decay) of line width is more rapid than the H$\alpha$ EW, which is often observed in M-dwarf flares \citep[e.g.,][]{2020PASJ...72...68N}. See Section \ref{sec:4-2} and \ref{sec:4-3}.
\item[(v)] The time evolution of the redshift velocity has a good agreement with Sun-as-a-star \textcolor{black}{near-ultraviolet} luminosity at 1600 {\AA} rather than optical white-light luminosity.
Also, the time evolution of the line width is similar to optical white lights, rather than H$\alpha$ EW. As for the latter, similar features have been reported in white-light and H$\alpha$ flares on M-dwarfs \citep[e.g.,][]{2020PASJ...72...68N}. The near-ultraviolet behaviors could be tested in future stellar observations, e.g., by using Swift/UVOT \citep[cf.,][]{2010ApJ...721..785O} and XMM-Newton \citep[cf.,][]{2005A&A...431..679M}. See Section \ref{sec:4-2} and \ref{sec:4-3}. 
\end{itemize}
These results of Sun-as-a-star analysis are useful in interpreting red asymmetries, as well as line broadenings, of the observed stellar flare spectra.  
The Sun-as-a-star values given in (i) and (iii) can be compared quantitatively with stellar flares \citep[e.g.,][]{2017ApJ...837..125K,2020PASJ...72...68N,2022ApJ...928..180W}, although it should be noted that the point (ii) is still subject for interpreting stellar flares. 
Also, if the time variation of redshift of observed stellar flares is the same as that of the result (iv)-(v), the line asymmetry is likely to come from flare ribbon radiation, while if it is not, it suggests the possibility of other sources of redshift radiation, such as flare loops or prominence activations \citep[cf.,][]{2022ApJ...928..180W}. 
\textcolor{black}{Our result is directly applicable to stars whose background H$\alpha$ lines shows absorption (e.g., G-type stars). 
Note that since pre-flare-subtracted spectrum is used in this study, the following caution should be taken for stars whose  background H$\alpha$ lines is emission (e.g., active M-type stars): the discussion of the qualitative nature of line asymmetry/broadening (velocity and line width) is applicable, but the ratio of line center intensity to red asymmetry intensity can be different (Section \ref{sec:4-2}).}

Sun-as-a-star analysis of H$\alpha$ line profiles of solar flares has been firstly conducted in \cite{Namekata2020Sci} for two events: a C-class solar flare associated with a dominated filament eruption and a C-class flare associated with a solar surge (i.e., a jet-like filament eruption).
In this paper, we have added an M-class solar flare with dominated flare-ribbons radiation as a new example.
The three examples show different behaviors in H$\alpha$ line, which supports a possibility that different phenomena on spatially-unresolved stars can be distinguished through H$\alpha$ line observations.
In the future work, the Sun-as-a-star analysis will be applied to a wider variety of phenomena including flare loops and prominence eruptions outside the limb as other patterns of solar phenomena (Otsu et al. in prep.).




\appendix

\color{black}

\section{Near-ultraviolet light curve and GOES time derivative}\label{sec:app-1}

\begin{figure}
\epsscale{0.5}
\plotone{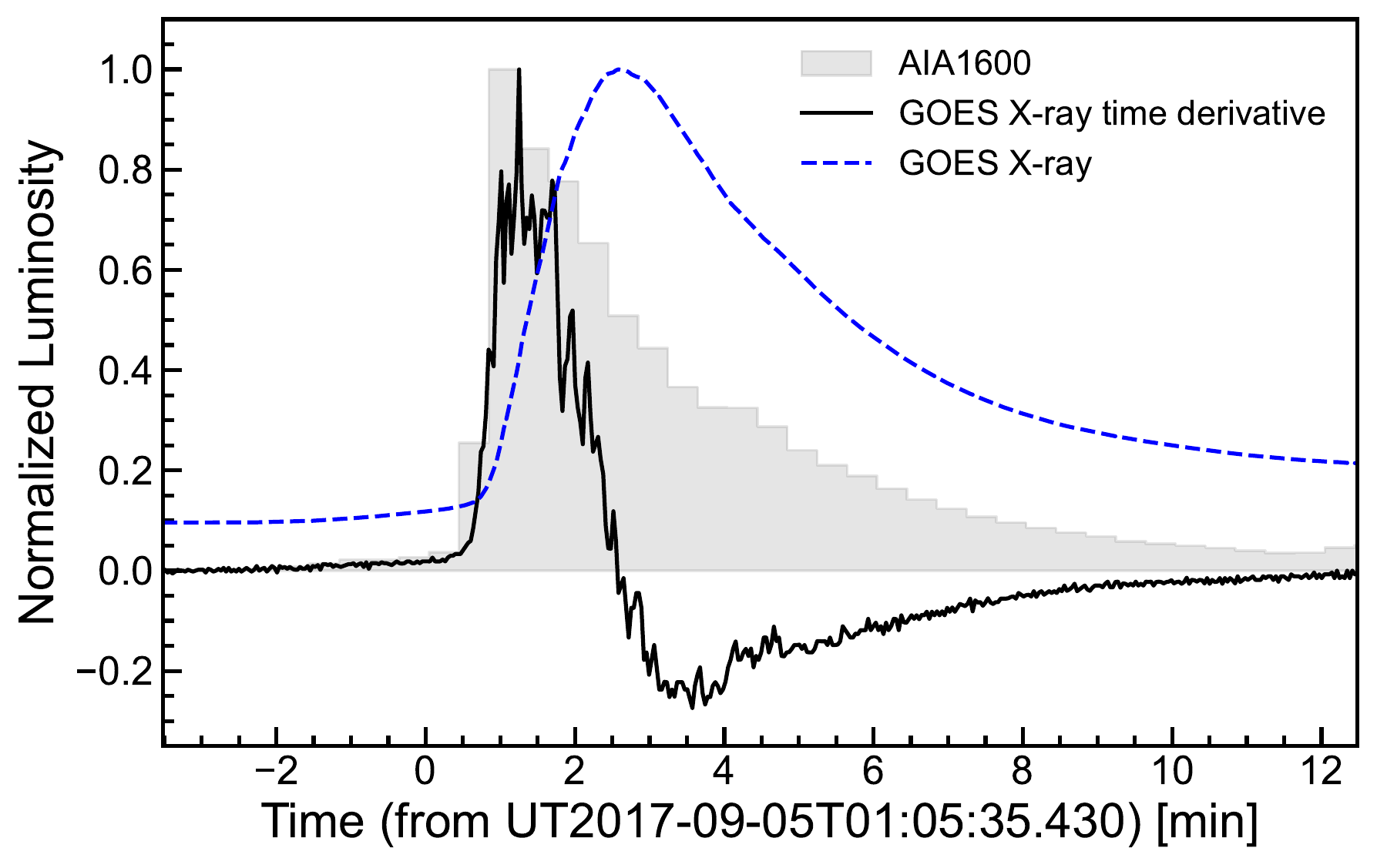}
\caption{\color{black} Comparison between 1600 {\AA} light curve (filled gray) and GOES time derivative (black). The original GOES soft X-ray light curve is also plotted with a blue dashed line.
}
\label{fig:13}
\end{figure}

Here we show a comparison between 1600 {\AA} light curve and GOES time derivative in Figure \ref{fig:13}. The peak timing of GOES time derivative is almost consistent with that of 1600 {\AA} light curve with an error of 10 second, although the decay timescale is different. This result is consistent with previous studies \citep{2009ApJ...694L..74N,2013ApJ...774...14Q,2018ApJ...863..124D}.

\color{black}

\begin{acknowledgments}
\textcolor{black}{The authors wish to thank the anonymous referee for helpful and constructive comments.} This research is supported by JSPS KAKENHI grant numbers 18J20048, 21J00316 (K.N.), and 21H01131 (K.S. and K.I.)
\end{acknowledgments}

%

\vspace{5mm}


\software{\textsf{astropy} \citep{2018AJ....156..123A}}

\bibliography{sample631,cvs}{}

\newcommand{\noop}[1]{}
\begin{thebibliography}{}
\expandafter\ifx\csname natexlab\endcsname\relax\def\natexlab#1{#1}\fi
\providecommand{\url}[1]{\href{#1}{#1}}
\providecommand{\dodoi}[1]{doi:~\href{http://doi.org/#1}{\nolinkurl{#1}}}
\providecommand{\doeprint}[1]{\href{http://ascl.net/#1}{\nolinkurl{http://ascl.net/#1}}}
\providecommand{\doarXiv}[1]{\href{https://arxiv.org/abs/#1}{\nolinkurl{https://arxiv.org/abs/#1}}}

\bibitem[{{Airapetian} {et~al.}(2016){Airapetian}, {Glocer}, {Gronoff},
  {H{\'e}brard}, \& {Danchi}}]{2016NatGe...9..452A}
{Airapetian}, V.~S., {Glocer}, A., {Gronoff}, G., {H{\'e}brard}, E., \&
  {Danchi}, W. 2016, Nature Geoscience, 9, 452, \dodoi{10.1038/ngeo2719}

\bibitem[{{Airapetian} {et~al.}(2020){Airapetian}, {Barnes}, {Cohen},
  {Collinson}, {Danchi}, {Dong}, {Del Genio}, {France}, {Garcia-Sage},
  {Glocer}, {Gopalswamy}, {Grenfell}, {Gronoff}, {G{\"u}del}, {Herbst},
  {Henning}, {Jackman}, {Jin}, {Johnstone}, {Kaltenegger}, {Kay}, {Kobayashi},
  {Kuang}, {Li}, {Lynch}, {L{\"u}ftinger}, {Luhmann}, {Maehara}, {Mlynczak},
  {Notsu}, {Osten}, {Ramirez}, {Rugheimer}, {Scheucher}, {Schlieder},
  {Shibata}, {Sousa-Silva}, {Stamenkovi{\'c}}, {Strangeway}, {Usmanov},
  {Vergados}, {Verkhoglyadova}, {Vidotto}, {Voytek}, {Way}, {Zank}, \&
  {Yamashiki}}]{2020IJAsB..19..136A}
{Airapetian}, V.~S., {Barnes}, R., {Cohen}, O., {et~al.} 2020, International
  Journal of Astrobiology, 19, 136, \dodoi{10.1017/S1473550419000132}

\bibitem[{{Asai} {et~al.}(2012){Asai}, {Ichimoto}, {Kita}, {Kurokawa}, \&
  {Shibata}}]{2012PASJ...64...20A}
{Asai}, A., {Ichimoto}, K., {Kita}, R., {Kurokawa}, H., \& {Shibata}, K. 2012,
  \pasj, 64, 20, \dodoi{10.1093/pasj/64.1.20}

\bibitem[{{Astropy Collaboration} {et~al.}(2018){Astropy Collaboration},
  {Price-Whelan}, {Sip{\H{o}}cz}, {G{\"u}nther}, {Lim}, {Crawford}, {Conseil},
  {Shupe}, {Craig}, {Dencheva}, {Ginsburg}, {VanderPlas}, {Bradley},
  {P{\'e}rez-Su{\'a}rez}, {de Val-Borro}, {Aldcroft}, {Cruz}, {Robitaille},
  {Tollerud}, {Ardelean}, {Babej}, {Bach}, {Bachetti}, {Bakanov}, {Bamford},
  {Barentsen}, {Barmby}, {Baumbach}, {Berry}, {Biscani}, {Boquien}, {Bostroem},
  {Bouma}, {Brammer}, {Bray}, {Breytenbach}, {Buddelmeijer}, {Burke},
  {Calderone}, {Cano Rodr{\'\i}guez}, {Cara}, {Cardoso}, {Cheedella}, {Copin},
  {Corrales}, {Crichton}, {D'Avella}, {Deil}, {Depagne}, {Dietrich}, {Donath},
  {Droettboom}, {Earl}, {Erben}, {Fabbro}, {Ferreira}, {Finethy}, {Fox},
  {Garrison}, {Gibbons}, {Goldstein}, {Gommers}, {Greco}, {Greenfield},
  {Groener}, {Grollier}, {Hagen}, {Hirst}, {Homeier}, {Horton}, {Hosseinzadeh},
  {Hu}, {Hunkeler}, {Ivezi{\'c}}, {Jain}, {Jenness}, {Kanarek}, {Kendrew},
  {Kern}, {Kerzendorf}, {Khvalko}, {King}, {Kirkby}, {Kulkarni}, {Kumar},
  {Lee}, {Lenz}, {Littlefair}, {Ma}, {Macleod}, {Mastropietro}, {McCully},
  {Montagnac}, {Morris}, {Mueller}, {Mumford}, {Muna}, {Murphy}, {Nelson},
  {Nguyen}, {Ninan}, {N{\"o}the}, {Ogaz}, {Oh}, {Parejko}, {Parley}, {Pascual},
  {Patil}, {Patil}, {Plunkett}, {Prochaska}, {Rastogi}, {Reddy Janga},
  {Sabater}, {Sakurikar}, {Seifert}, {Sherbert}, {Sherwood-Taylor}, {Shih},
  {Sick}, {Silbiger}, {Singanamalla}, {Singer}, {Sladen}, {Sooley},
  {Sornarajah}, {Streicher}, {Teuben}, {Thomas}, {Tremblay}, {Turner},
  {Terr{\'o}n}, {van Kerkwijk}, {de la Vega}, {Watkins}, {Weaver}, {Whitmore},
  {Woillez}, {Zabalza}, \& {Astropy Contributors}}]{2018AJ....156..123A}
{Astropy Collaboration}, {Price-Whelan}, A.~M., {Sip{\H{o}}cz}, B.~M., {et~al.}
  2018, \aj, 156, 123, \dodoi{10.3847/1538-3881/aabc4f}

\bibitem[{{Beckers}(1964)}]{1964PhDT........83B}
{Beckers}, J.~M. 1964, PhD thesis, Sacramento Peak Observatory, Air Force
  Cambridge Research Laboratories, Mass., USA

\bibitem[{{Canfield} {et~al.}(1990){Canfield}, {Penn}, {Wulser}, \&
  {Kiplinger}}]{1990ApJ...363..318C}
{Canfield}, R.~C., {Penn}, M.~J., {Wulser}, J.-P., \& {Kiplinger}, A.~L. 1990,
  \apj, 363, 318, \dodoi{10.1086/169345}

\bibitem[{{Chertok} {et~al.}(2018){Chertok}, {Belov}, \&
  {Abunin}}]{2018SpWea..16.1549C}
{Chertok}, I.~M., {Belov}, A.~V., \& {Abunin}, A.~A. 2018, Space Weather, 16,
  1549, \dodoi{10.1029/2018SW001899}

\bibitem[{{Dai} {et~al.}(2018){Dai}, {Ding}, {Zong}, \&
  {Yang}}]{2018ApJ...863..124D}
{Dai}, Y., {Ding}, M., {Zong}, W., \& {Yang}, K.~E. 2018, \apj, 863, 124,
  \dodoi{10.3847/1538-4357/aad32e}

\bibitem[{{Fisher} {et~al.}(1985){Fisher}, {Canfield}, \&
  {McClymont}}]{1985ApJ...289..414F}
{Fisher}, G.~H., {Canfield}, R.~C., \& {McClymont}, A.~N. 1985, \apj, 289, 414,
  \dodoi{10.1086/162901}

\bibitem[{{Fletcher} \& {Hudson}(2008)}]{2008ApJ...675.1645F}
{Fletcher}, L., \& {Hudson}, H.~S. 2008, \apj, 675, 1645,
  \dodoi{10.1086/527044}

\bibitem[{{Fuhrmeister} {et~al.}(2011){Fuhrmeister}, {Lalitha}, {Poppenhaeger},
  {Rudolf}, {Liefke}, {Reiners}, {Schmitt}, \& {Ness}}]{2011A&A...534A.133F}
{Fuhrmeister}, B., {Lalitha}, S., {Poppenhaeger}, K., {et~al.} 2011, \aap, 534,
  A133, \dodoi{10.1051/0004-6361/201117447}

\bibitem[{{Fuhrmeister} \& {Schmitt}(2004)}]{2004A&A...420.1079F}
{Fuhrmeister}, B., \& {Schmitt}, J.~H.~M.~M. 2004, \aap, 420, 1079,
  \dodoi{10.1051/0004-6361:20035644}

\bibitem[{{Guarcello} {et~al.}(2019){Guarcello}, {Micela}, {Sciortino},
  {L{\'o}pez-Santiago}, {Argiroffi}, {Reale}, {Flaccomio},
  {Alvarado-G{\'o}mez}, {Antoniou}, {Drake}, {Pillitteri}, {Rebull}, \&
  {Stauffer}}]{2019A&A...622A.210G}
{Guarcello}, M.~G., {Micela}, G., {Sciortino}, S., {et~al.} 2019, \aap, 622,
  A210, \dodoi{10.1051/0004-6361/201834370}

\bibitem[{{Gunn} {et~al.}(1994){Gunn}, {Doyle}, {Mathioudakis}, {Houdebine}, \&
  {Avgoloupis}}]{1994A&A...285..489G}
{Gunn}, A.~G., {Doyle}, J.~G., {Mathioudakis}, M., {Houdebine}, E.~R., \&
  {Avgoloupis}, S. 1994, \aap, 285, 489

\bibitem[{{Harra} {et~al.}(2016){Harra}, {Schrijver}, {Janvier}, {Toriumi},
  {Hudson}, {Matthews}, {Woods}, {Hara}, {Guedel}, {Kowalski}, {Osten},
  {Kusano}, \& {Lueftinger}}]{2016SoPh..291.1761H}
{Harra}, L.~K., {Schrijver}, C.~J., {Janvier}, M., {et~al.} 2016, \solphys,
  291, 1761, \dodoi{10.1007/s11207-016-0923-0}

\bibitem[{{Honda} {et~al.}(2018){Honda}, {Notsu}, {Namekata}, {Notsu},
  {Maehara}, {Ikuta}, {Nogami}, \& {Shibata}}]{2018PASJ...70...62H}
{Honda}, S., {Notsu}, Y., {Namekata}, K., {et~al.} 2018, \pasj, 70, 62,
  \dodoi{10.1093/pasj/psy055}

\bibitem[{{Houdebine} {et~al.}(1993){Houdebine}, {Foing}, {Doyle}, \&
  {Rodono}}]{1993A&A...274..245H}
{Houdebine}, E.~R., {Foing}, B.~H., {Doyle}, J.~G., \& {Rodono}, M. 1993, \aap,
  274, 245

\bibitem[{{Houdebine} {et~al.}(1990){Houdebine}, {Foing}, \&
  {Rodono}}]{1990A&A...238..249H}
{Houdebine}, E.~R., {Foing}, B.~H., \& {Rodono}, M. 1990, \aap, 238, 249

\bibitem[{{Ichimoto} \& {Kurokawa}(1984)}]{1984SoPh...93..105I}
{Ichimoto}, K., \& {Kurokawa}, H. 1984, \solphys, 93, 105,
  \dodoi{10.1007/BF00156656}

\bibitem[{{Ichimoto} {et~al.}(2017){Ichimoto}, {Ishii}, {Otsuji}, {Kimura},
  {Nakatani}, {Kaneda}, {Nagata}, {UeNo}, {Hirose}, {Cabezas}, \&
  {Morita}}]{2017SoPh..292...63I}
{Ichimoto}, K., {Ishii}, T.~T., {Otsuji}, K., {et~al.} 2017, \solphys, 292, 63,
  \dodoi{10.1007/s11207-017-1082-7}

\bibitem[{{Jing} {et~al.}(2016){Jing}, {Xu}, {Cao}, {Liu}, {Gary}, \&
  {Wang}}]{2016NatSR...624319J}
{Jing}, J., {Xu}, Y., {Cao}, W., {et~al.} 2016, Scientific Reports, 6, 24319,
  \dodoi{10.1038/srep24319}

\bibitem[{{Koller} {et~al.}(2020){Koller}, {Leitzinger}, {Temmer}, {Odert},
  {Beck}, \& {Veronig}}]{2020arXiv201200786K}
{Koller}, F., {Leitzinger}, M., {Temmer}, M., {et~al.} 2020, arXiv e-prints,
  arXiv:2012.00786.
\newblock \doarXiv{2012.00786}

\bibitem[{{Kowalski} {et~al.}(2022){Kowalski}, {Allred}, {Carlsson}, {Kerr},
  {Tremblay}, {Namekata}, {Kuridze}, \& {Uitenbroek}}]{2022arXiv220113349K}
{Kowalski}, A.~F., {Allred}, J.~C., {Carlsson}, M., {et~al.} 2022, arXiv
  e-prints, arXiv:2201.13349.
\newblock \doarXiv{2201.13349}

\bibitem[{{Kowalski} {et~al.}(2015){Kowalski}, {Hawley}, {Carlsson}, {Allred},
  {Uitenbroek}, {Osten}, \& {Holman}}]{2015SoPh..290.3487K}
{Kowalski}, A.~F., {Hawley}, S.~L., {Carlsson}, M., {et~al.} 2015, \solphys,
  290, 3487, \dodoi{10.1007/s11207-015-0708-x}

\bibitem[{{Kowalski} {et~al.}(2017){Kowalski}, {Allred}, {Uitenbroek},
  {Tremblay}, {Brown}, {Carlsson}, {Osten}, {Wisniewski}, \&
  {Hawley}}]{2017ApJ...837..125K}
{Kowalski}, A.~F., {Allred}, J.~C., {Uitenbroek}, H., {et~al.} 2017, \apj, 837,
  125, \dodoi{10.3847/1538-4357/aa603e}

\bibitem[{{Kretzschmar}(2011)}]{2011A&A...530A..84K}
{Kretzschmar}, M. 2011, \aap, 530, A84, \dodoi{10.1051/0004-6361/201015930}

\bibitem[{{Kretzschmar} {et~al.}(2010){Kretzschmar}, {de Wit}, {Schmutz},
  {Mekaoui}, {Hochedez}, \& {Dewitte}}]{2010NatPh...6..690K}
{Kretzschmar}, M., {de Wit}, T.~D., {Schmutz}, W., {et~al.} 2010, Nature
  Physics, 6, 690, \dodoi{10.1038/nphys1741}

\bibitem[{{Livshits} {et~al.}(1981){Livshits}, {Badalian}, {Kosovichev}, \&
  {Katsova}}]{1981SoPh...73..269L}
{Livshits}, M.~A., {Badalian}, O.~G., {Kosovichev}, A.~G., \& {Katsova}, M.~M.
  1981, \solphys, 73, 269, \dodoi{10.1007/BF00151682}

\bibitem[{{Maehara} {et~al.}(2012){Maehara}, {Shibayama}, {Notsu}, {Notsu},
  {Nagao}, {Kusaba}, {Honda}, {Nogami}, \& {Shibata}}]{2012Natur.485..478M}
{Maehara}, H., {Shibayama}, T., {Notsu}, S., {et~al.} 2012, Nature, 485, 478,
  \dodoi{10.1038/nature11063}

\bibitem[{{Maehara} {et~al.}(2021){Maehara}, {Notsu}, {Namekata}, {Honda},
  {Kowalski}, {Katoh}, {Ohshima}, {Iida}, {Oeda}, {Murata}, {Yamanaka},
  {Takagi}, {Sasada}, {Akitaya}, {Ikuta}, {Okamoto}, {Nogami}, \&
  {Shibata}}]{2020PASJ..tmp..253M}
{Maehara}, H., {Notsu}, Y., {Namekata}, K., {et~al.} 2021, \pasj,
  \dodoi{10.1093/pasj/psaa098}

\bibitem[{{Mitra-Kraev} {et~al.}(2005){Mitra-Kraev}, {Harra}, {G{\"u}del},
  {Audard}, {Branduardi-Raymont}, {Kay}, {Mewe}, {Raassen}, \& {van
  Driel-Gesztelyi}}]{2005A&A...431..679M}
{Mitra-Kraev}, U., {Harra}, L.~K., {G{\"u}del}, M., {et~al.} 2005, \aap, 431,
  679, \dodoi{10.1051/0004-6361:20041201}

\bibitem[{{Muheki} {et~al.}(2020){Muheki}, {Guenther}, {Mutabazi}, \&
  {Jurua}}]{2020MNRAS.499.5047M}
{Muheki}, P., {Guenther}, E.~W., {Mutabazi}, T., \& {Jurua}, E. 2020, \mnras,
  499, 5047, \dodoi{10.1093/mnras/staa3152}

\bibitem[{{Namekata} {et~al.}(2017{\natexlab{a}}){Namekata}, {Sakaue},
  {Watanabe}, {Asai}, \& {Shibata}}]{2017PASJ...69....7N}
{Namekata}, K., {Sakaue}, T., {Watanabe}, K., {Asai}, A., \& {Shibata}, K.
  2017{\natexlab{a}}, \pasj, 69, 7, \dodoi{10.1093/pasj/psw111}

\bibitem[{{Namekata} {et~al.}(2017{\natexlab{b}}){Namekata}, {Sakaue},
  {Watanabe}, {Asai}, {Maehara}, {Notsu}, {Notsu}, {Honda}, {Ishii}, {Ikuta},
  {Nogami}, \& {Shibata}}]{2017ApJ...851...91N}
{Namekata}, K., {Sakaue}, T., {Watanabe}, K., {et~al.} 2017{\natexlab{b}},
  \apj, 851, 91, \dodoi{10.3847/1538-4357/aa9b34}

\bibitem[{{Namekata} {et~al.}(2020){Namekata}, {Maehara}, {Sasaki}, {Kawai},
  {Notsu}, {Kowalski}, {Allred}, {Iwakiri}, {Tsuboi}, {Murata}, {Niwano},
  {Shiraishi}, {Adachi}, {Iida}, {Oeda}, {Honda}, {Tozuka}, {Katoh}, {Onozato},
  {Okamoto}, {Isogai}, {Kimura}, {Kojiguchi}, {Wakamatsu}, {Tampo}, {Nogami},
  \& {Shibata}}]{2020PASJ...72...68N}
{Namekata}, K., {Maehara}, H., {Sasaki}, R., {et~al.} 2020, \pasj, 72, 68,
  \dodoi{10.1093/pasj/psaa051}

\bibitem[{{Namekata} {et~al.}(2022{\natexlab{a}}){Namekata}, {Maehara},
  {Honda}, {Notsu}, {Okamoto}, {Takahashi}, {Takayama}, {Ohshima}, {Saito},
  {Katoh}, {Tozuka}, {Murata}, {Ogawa}, {Niwano}, {Adachi}, {Oeda},
  {Shiraishi}, {Isogai}, {Seki}, {Ishii}, {Ichimoto}, {Nogami}, \&
  {Shibata}}]{Namekata2020Sci}
{Namekata}, K., {Maehara}, H., {Honda}, S., {et~al.} 2022{\natexlab{a}}, Nature
  Astronomy, 6, 241, \dodoi{10.1038/s41550-021-01532-8}

\bibitem[{{Namekata} {et~al.}(2022{\natexlab{b}}){Namekata}, {Maehara},
  {Honda}, {Notsu}, {Okamoto}, {Takahashi}, {Takayama}, {Ohshima}, {Saito},
  {Katoh}, {Tozuka}, {Murata}, {Ogawa}, {Niwano}, {Adachi}, {Oeda},
  {Shiraishi}, {Isogai}, {Nogami}, \& {Shibata}}]{2022arXiv220109416N}
---. 2022{\natexlab{b}}, \apjl, 926, L5, \dodoi{10.3847/2041-8213/ac4df0}

\bibitem[{{Nishizuka} {et~al.}(2009){Nishizuka}, {Asai}, {Takasaki},
  {Kurokawa}, \& {Shibata}}]{2009ApJ...694L..74N}
{Nishizuka}, N., {Asai}, A., {Takasaki}, H., {Kurokawa}, H., \& {Shibata}, K.
  2009, \apjl, 694, L74, \dodoi{10.1088/0004-637X/694/1/L74}

\bibitem[{{Notsu} {et~al.}(2019){Notsu}, {Maehara}, {Honda}, {Hawley},
  {Davenport}, {Namekata}, {Notsu}, {Ikuta}, {Nogami}, \&
  {Shibata}}]{2019ApJ...876...58N}
{Notsu}, Y., {Maehara}, H., {Honda}, S., {et~al.} 2019, \apj, 876, 58,
  \dodoi{10.3847/1538-4357/ab14e6}

\bibitem[{{Okamoto} {et~al.}(2021){Okamoto}, {Notsu}, {Maehara}, {Namekata},
  {Honda}, {Ikuta}, {Nogami}, \& {Shibata}}]{2020arXiv201102117O}
{Okamoto}, S., {Notsu}, Y., {Maehara}, H., {et~al.} 2021, \apj, 906, 72,
  \dodoi{10.3847/1538-4357/abc8f5}

\bibitem[{{Osten} \& {Wolk}(2015)}]{2015ApJ...809...79O}
{Osten}, R.~A., \& {Wolk}, S.~J. 2015, \apj, 809, 79,
  \dodoi{10.1088/0004-637X/809/1/79}

\bibitem[{{Osten} {et~al.}(2010){Osten}, {Godet}, {Drake}, {Tueller},
  {Cummings}, {Krimm}, {Pye}, {Pal'shin}, {Golenetskii}, {Reale}, {Oates},
  {Page}, \& {Melandri}}]{2010ApJ...721..785O}
{Osten}, R.~A., {Godet}, O., {Drake}, S., {et~al.} 2010, \apj, 721, 785,
  \dodoi{10.1088/0004-637X/721/1/785}

\bibitem[{{Parenti}(2014)}]{2014LRSP...11....1P}
{Parenti}, S. 2014, Living Reviews in Solar Physics, 11, 1,
  \dodoi{10.12942/lrsp-2014-1}

\bibitem[{{Priest}(1981)}]{1981sfmh.book.....P}
{Priest}, E.~R. 1981, {Solar flare magnetohydrodynamics}

\bibitem[{{Qiu} {et~al.}(2013){Qiu}, {Sturrock}, {Longcope}, {Klimchuk}, \&
  {Liu}}]{2013ApJ...774...14Q}
{Qiu}, J., {Sturrock}, Z., {Longcope}, D.~W., {Klimchuk}, J.~A., \& {Liu},
  W.-J. 2013, \apj, 774, 14, \dodoi{10.1088/0004-637X/774/1/14}

\bibitem[{{Reale} {et~al.}(1997){Reale}, {Betta}, {Peres}, {Serio}, \&
  {McTiernan}}]{1997A&A...325..782R}
{Reale}, F., {Betta}, R., {Peres}, G., {Serio}, S., \& {McTiernan}, J. 1997,
  \aap, 325, 782

\bibitem[{{Schrijver} {et~al.}(2012){Schrijver}, {Beer}, {Baltensperger},
  {Cliver}, {G{\"u}del}, {Hudson}, {McCracken}, {Osten}, {Peter}, {Soderblom},
  {Usoskin}, \& {Wolff}}]{2012JGRA..117.8103S}
{Schrijver}, C.~J., {Beer}, J., {Baltensperger}, U., {et~al.} 2012, Journal of
  Geophysical Research (Space Physics), 117, A08103,
  \dodoi{10.1029/2012JA017706}

\bibitem[{{Segura} {et~al.}(2010){Segura}, {Walkowicz}, {Meadows}, {Kasting},
  \& {Hawley}}]{2010AsBio..10..751S}
{Segura}, A., {Walkowicz}, L.~M., {Meadows}, V., {Kasting}, J., \& {Hawley}, S.
  2010, Astrobiology, 10, 751, \dodoi{10.1089/ast.2009.0376}

\bibitem[{{Seki} {et~al.}(2017){Seki}, {Otsuji}, {Isobe}, {Ishii}, {Sakaue}, \&
  {Hirose}}]{2017ApJ...843L..24S}
{Seki}, D., {Otsuji}, K., {Isobe}, H., {et~al.} 2017, \apjl, 843, L24,
  \dodoi{10.3847/2041-8213/aa7559}

\bibitem[{{Shen} {et~al.}(2018){Shen}, {Xu}, {Wang}, {Chi}, \&
  {Luo}}]{2018ApJ...861...28S}
{Shen}, C., {Xu}, M., {Wang}, Y., {Chi}, Y., \& {Luo}, B. 2018, \apj, 861, 28,
  \dodoi{10.3847/1538-4357/aac204}

\bibitem[{{Shibata} \& {Magara}(2011)}]{2011LRSP....8....6S}
{Shibata}, K., \& {Magara}, T. 2011, Living Reviews in Solar Physics, 8, 6,
  \dodoi{10.12942/lrsp-2011-6}

\bibitem[{{Shibata} \& {Yokoyama}(2002)}]{2002ApJ...577..422S}
{Shibata}, K., \& {Yokoyama}, T. 2002, \apj, 577, 422, \dodoi{10.1086/342141}

\bibitem[{{Shibata} {et~al.}(2013){Shibata}, {Isobe}, {Hillier}, {Choudhuri},
  {Maehara}, {Ishii}, {Shibayama}, {Notsu}, {Notsu}, {Nagao}, {Honda}, \&
  {Nogami}}]{2013PASJ...65...49S}
{Shibata}, K., {Isobe}, H., {Hillier}, A., {et~al.} 2013, \pasj, 65, 49,
  \dodoi{10.1093/pasj/65.3.49}

\bibitem[{{Shibayama} {et~al.}(2013){Shibayama}, {Maehara}, {Notsu}, {Notsu},
  {Nagao}, {Honda}, {Ishii}, {Nogami}, \& {Shibata}}]{2013ApJS..209....5S}
{Shibayama}, T., {Maehara}, H., {Notsu}, S., {et~al.} 2013, \apjs, 209, 5,
  \dodoi{10.1088/0067-0049/209/1/5}

\bibitem[{{Sim{\~o}es} {et~al.}(2019){Sim{\~o}es}, {Reid}, {Milligan}, \&
  {Fletcher}}]{2019ApJ...870..114S}
{Sim{\~o}es}, P. J.~A., {Reid}, H. A.~S., {Milligan}, R.~O., \& {Fletcher}, L.
  2019, \apj, 870, 114, \dodoi{10.3847/1538-4357/aaf28d}

\bibitem[{{Somov} {et~al.}(1982){Somov}, {Sermulina}, \&
  {Spektor}}]{1982SoPh...81..281S}
{Somov}, B.~V., {Sermulina}, B.~J., \& {Spektor}, A.~R. 1982, \solphys, 81,
  281, \dodoi{10.1007/BF00151302}

\bibitem[{{Soni} {et~al.}(2020){Soni}, {Gupta}, \&
  {Verma}}]{2020RAA....20...23S}
{Soni}, S.~L., {Gupta}, R.~S., \& {Verma}, P.~L. 2020, Research in Astronomy
  and Astrophysics, 20, 023, \dodoi{10.1088/1674-4527/20/2/23}

\bibitem[{{Tei} {et~al.}(2018){Tei}, {Sakaue}, {Okamoto}, {Kawate}, {Heinzel},
  {UeNo}, {Asai}, {Ichimoto}, \& {Shibata}}]{2018PASJ...70..100T}
{Tei}, A., {Sakaue}, T., {Okamoto}, T.~J., {et~al.} 2018, \pasj, 70, 100,
  \dodoi{10.1093/pasj/psy047}

\bibitem[{{Temmer}(2021)}]{2021LRSP...18....4T}
{Temmer}, M. 2021, Living Reviews in Solar Physics, 18, 4,
  \dodoi{10.1007/s41116-021-00030-3}

\bibitem[{{Veronig} {et~al.}(2021){Veronig}, {Odert}, {Leitzinger}, {Dissauer},
  {Fleck}, \& {Hudson}}]{2021NatAs...5..697V}
{Veronig}, A.~M., {Odert}, P., {Leitzinger}, M., {et~al.} 2021, Nature
  Astronomy, 5, 697, \dodoi{10.1038/s41550-021-01345-9}

\bibitem[{{Vida} {et~al.}(2019){Vida}, {Leitzinger}, {Kriskovics}, {Seli},
  {Odert}, {Kov{\'a}cs}, {Korhonen}, \& {van
  Driel-Gesztelyi}}]{2019A&A...623A..49V}
{Vida}, K., {Leitzinger}, M., {Kriskovics}, L., {et~al.} 2019, \aap, 623, A49,
  \dodoi{10.1051/0004-6361/201834264}

\bibitem[{{Vida} {et~al.}(2016){Vida}, {Kriskovics}, {Ol{\'a}h}, {Leitzinger},
  {Odert}, {K{\H{o}}v{\'a}ri}, {Korhonen}, {Greimel}, {Robb}, {Cs{\'a}k}, \&
  {Kov{\'a}cs}}]{2016A&A...590A..11V}
{Vida}, K., {Kriskovics}, L., {Ol{\'a}h}, K., {et~al.} 2016, \aap, 590, A11,
  \dodoi{10.1051/0004-6361/201527925}

\bibitem[{{{\v{S}}vestka} {et~al.}(1962){{\v{S}}vestka}, {Kopeck{\'y}}, \&
  {Blaha}}]{1962BAICz..13...37S}
{{\v{S}}vestka}, Z., {Kopeck{\'y}}, M., \& {Blaha}, M. 1962, Bulletin of the
  Astronomical Institutes of Czechoslovakia, 13, 37

\bibitem[{{Wiik} {et~al.}(1996){Wiik}, {Schmieder}, {Heinzel}, \&
  {Roudier}}]{1996SoPh..166...89W}
{Wiik}, J.~E., {Schmieder}, B., {Heinzel}, P., \& {Roudier}, T. 1996, \solphys,
  166, 89, \dodoi{10.1007/BF00179357}

\bibitem[{{Wu} {et~al.}(2022){Wu}, {Chen}, {Tian}, {Zhang}, {Shi}, {He}, {Lu},
  {Xu}, \& {Wang}}]{2022ApJ...928..180W}
{Wu}, Y., {Chen}, H., {Tian}, H., {et~al.} 2022, \apj, 928, 180,
  \dodoi{10.3847/1538-4357/ac5897}

\bibitem[{{Yamasaki} {et~al.}(2021){Yamasaki}, {Inoue}, {Nagata}, \&
  {Ichimoto}}]{2021ApJ...908..132Y}
{Yamasaki}, D., {Inoue}, S., {Nagata}, S., \& {Ichimoto}, K. 2021, \apj, 908,
  132, \dodoi{10.3847/1538-4357/abcfbb}

\bibitem[{{Yamashiki} {et~al.}(2019){Yamashiki}, {Maehara}, {Airapetian},
  {Notsu}, {Sato}, {Notsu}, {Kuroki}, {Murashima}, {Sato}, {Namekata},
  {Sasaki}, {Scott}, {Bando}, {Nashimoto}, {Takagi}, {Ling}, {Nogami}, \&
  {Shibata}}]{2019ApJ...881..114Y}
{Yamashiki}, Y.~A., {Maehara}, H., {Airapetian}, V., {et~al.} 2019, \apj, 881,
  114, \dodoi{10.3847/1538-4357/ab2a71}

\bibitem[{{Yan} {et~al.}(2018){Yan}, {Wang}, {Pan}, {Kong}, {Xue}, {Yang},
  {Li}, \& {Feng}}]{2018ApJ...856...79Y}
{Yan}, X.~L., {Wang}, J.~C., {Pan}, G.~M., {et~al.} 2018, \apj, 856, 79,
  \dodoi{10.3847/1538-4357/aab153}

\bibitem[{{Zhou} {et~al.}(2020){Zhou}, {Li}, {Ding}, {Hong}, \&
  {Yu}}]{2020ApJ...904...95Z}
{Zhou}, Y.-A., {Li}, Y., {Ding}, M.~D., {Hong}, J., \& {Yu}, K. 2020, \apj,
  904, 95, \dodoi{10.3847/1538-4357/abb77c}

\end{thebibliography}
\bibliographystyle{aasjournal}



\end{document}